\documentclass[aps,11pt,prd,groupedaddress,nofootinbib,notitlepage,eqsecnum,preprintnumbers]{revtex4-2}
\usepackage[utf8]{inputenc}
\usepackage{hyperref}
\usepackage[table]{xcolor}
\usepackage{graphicx}
\usepackage{amsmath,amssymb}
\usepackage{bm}
\usepackage{comment}
\usepackage[shortlabels]{enumitem}
\usepackage{overpic}
\usepackage{ulem}
\usepackage{array}
\usepackage{tabularx}
\usepackage{booktabs}
\usepackage{diagbox}

\begin{document}
\title{Galaxy number-count dipole and superhorizon fluctuations}

\author{\textsc{Guillem Dom\`enech$^{a}$}}
    \email{{domenech}@{pd.infn.it}}
\author{\textsc{Roya Mohayaee$^{b}$}}
\author{\textsc{Subodh P. Patil$^{c}$}}
\author{\textsc{Subir Sarkar$^{d}$}}

\affiliation{$^a$INFN Sezione di Padova, via Marzolo 8, I-35131 Padova, Italy}
\affiliation{$^b$ Sorbonne Université, CNRS, Institut d’Astrophysique de Paris (IAP), 98 bis Bld Arago, Paris, France}
\affiliation{$^{c}$ Instituut-Lorentz for Theoretical Physics, Leiden University, 2333 CA Leiden, Netherlands}
\affiliation{$^{d}$ Rudolf Peierls Centre for Theoretical Physics, University of Oxford, Parks Road, Oxford, OX1 3PU, United Kingdom}

\begin{abstract}
In view of the growing tension between the dipole anisotropy of number counts of cosmologically distant sources and of the cosmic microwave background (CMB), we investigate the number count dipole induced by primordial perturbations with wavelength comparable to or exceeding the Hubble radius today. First, we find that neither adiabatic nor isocurvature superhorizon modes can generate an intrinsic number count dipole. However a superhorizon isocurvature mode does induce a relative velocity between the CMB and the (dark) matter rest frames and thereby affects the CMB dipole. We revisit the possibility that it has an intrinsic component due to such a mode, thus enabling consistency with the galaxy number count dipole if the latter is actually kinematic in origin. Although this scenario is not particularly natural, there are possible links with other anomalies and it predicts a concommitant galaxy number count quadrupole which may be measurable in future surveys. We also investigate the number count dipole induced by modes smaller than the Hubble radius, finding that subject to CMB constraints this is too small to reconcile the dipole tension.
\end{abstract}

\maketitle

\section{Introduction \label{sec:intro}}


Despite great advances in cosmology, we have yet to confirm whether the cosmic rest frame (CRF) of galaxies and dark matter in which they are isotropically distributed on the sky, actually coincides with that of the Cosmic Microwave Background (CMB). Ellis \& Baldwin \cite{ellisbaldwin} were the first to note that this is a powerful consistency test of the Friedman-Lemaître-Robertson-Walker (FLRW) models, in which the two CRFs must coincide. If the observed CMB dipole anisotropy is indeed kinematic due to our local peculiar motion with respect to the CRF \cite{Sciama:1967zz}, then there must be a corresponding kinematic dipole imprinted in the galaxy number counts on the sky.
 
It is only recently that large enough galaxy samples have become available to perform this test \cite{Condon:1998iy,Mauch:2003zh,Intema:2016jhx,Eisenhardt_2020,Lacy:2019rfe,McConnell:2020obn}. Interestingly, there is growing evidence that the two CRFs do \textit{not} coincide \cite{Blake:2002gx,Singal:2011dy,Gibelyou:2012ri,Rubart:2013tx,Tiwari:2013vff,Tiwari:2013ima,Tiwari:2015tba,Colin:2017juj,Bengaly:2017slg,Singal:2019pqq,Siewert:2020krp,Secrest:2020has} thus challenging the fundamental basis of the standard $\Lambda$CDM cosmology. In fact distant radio galaxies and quasars exhibit a dipole 2--3 times higher than expected from the kinematic interpretation of CMB dipole, with the highest reported tension now exceeding $5\sigma$ \cite{Secrest:2020has,Secrest:2022uvx}.

The dipole test \cite{ellisbaldwin} is sometimes referred to as a test of the Cosmological Principle \cite{Secrest:2020has,Nadolny:2021hti}.\footnote{The Cosmological Principle asserts that the universe must appear to be the same to all Observers, wherever they are. It is then homogeneous \& isotropic so can be described by the FLRW metric. However we have a peculiar (non-Hubble) motion wrt the CRF in which the above holds, so all cosmological measurements we make must first be transformed to the CRF in order for the Friedman-Lemaître (FL) equations to be applicable.} However we must first rule out that the dipole mismatch is not due to any large-scale inhomogeneity compatible with a perturbed FLRW model before abandoning this foundational assumption of the standard cosmology, upon which the inference that the universe is dominated today by a Cosmological constant $\Lambda$ rests \cite{Sarkar:2007cx}. After all, apart from its dipole, the CMB has only small anisotropies of ${\cal O}(10^{-5})$ \cite{Planck:2018nkj,Planck:2019evm}, and peculiar velocities are of ${\cal O}(10^{-3}c)$ \textit{i.e.} well within the validity of cosmological perturbation theory. In fact it is quite possible that (dark) matter has a relative velocity with respect to the CMB, which is often a signature of isocurvature fluctuations \cite{Kodama:1986fg,Kodama:1986ud,Turner:1991di,Turner:1991dn,Ma:2010ps}. Moreover, the dipole tension appears only if we assume that there are no intrinsic (non-kinematic) components to either dipole. This is a strong assumption which requires scrutiny. While the kinematic interpretation of the CMB dipole is consistent with other tests \cite{Planck:2013kqc,Planck:2020qil,Saha:2021bay}, these do not exclude the possibility of an intrinsic dipole \cite{Roldan:2016ayx,Ferreira:2020aqa,Ferreira:2021omv}. This motivates our investigation of whether the dipole tension can be explained by inhomogeneities larger than the present horizon.

Turner \cite{Turner:1991dn,Turner:1991di} (see also Ref.~\cite{Gunn:1988}) suggested that a superhorizon isocurvature fluctuation can provide an intrinsic dipole modulation to the CMB and called it a ``tilted universe''\footnote{This should not be misinterpreted as meaning that the universe is not isotropic on average on the largest scales. Furthermore, what is ``tilted'' depends on the Observer --- e.g. in the (dark) matter rest frame galaxies are not going anywhere, but the CMB is sliding instead.} to express that it would look as if galaxies have a global motion with respect to the CMB rest frame. 
The effects of superhorizon fluctuations on the CMB have subsequently been studied in detail \cite{Langlois:1995ca,Zibin:2008fe,Erickcek:2008sm,Erickcek:2008jp,Erickcek:2009at} and it is established that only isocurvature modes induce a CMB dipole at leading order. In this work, we also study the effect on the galaxy number count dipole using the formalism developed in Refs.~\cite{Kasai:1987ap,Yoo:2009au,Bonvin:2011bg,Challinor:2011bk}.\footnote{The first application of cosmological perturbation theory to galaxy number counts was by Kasai \& Sasaki \cite{Kasai:1987ap} who also demonstrated invariance under gauge transformations.} Related work may be found in Refs.~\cite{Ghosh:2013blq,Das:2021ssc,Tiwari:2021ikr}.

It is important to explore such new physics also in view of other reported anomalies and tensions within $\Lambda$CDM (see e.g. Refs.~\cite{Schwarz:2015cma,Perivolaropoulos:2021jda,Abdalla:2022yfr} for recent reviews). For example, the effects of a local void on the CMB dipole have been discussed  ~\cite{1990ApJ...364..341P,Alnes:2006uk,Rubart:2014lia}. The presence of superhorizon modes is an interesting possibility in itself as they may be pre-inflationary remnants\footnote{Isocurvature superhorizon modes can also be excited in double inflation \cite{Langlois:1996ms}.} and for the mode  to not have been diluted away, inflation must have lasted for just long enough to create our present Hubble patch.
This scenario fits well within open inflation models wherein our observable universe results from bubble nucleation in false vacuum \cite{
Gott:1982zf,Langlois:1996rx,Garcia-Bellido:1997tjw,Linde:1998iw,Linde:1999wv}. Furthermore, isocurvature fluctuations typically arise in the presence of light scalar fields during inflation \cite{Axenides:1983hj,Linde:1985yf,Seckel:1985tj}.  This applies, for instance, to axions \cite{Preskill:1982cy,Abbott:1982af,Dine:1982ah,Svrcek:2006yi} (see, \textit{e.g.}, Refs.~\cite{Marsh:2015xka,Irastorza:2018dyq} for recent reviews of axion cosmology), or more generically to curvatons \cite{Lyth:2001nq,Enqvist:2001zp,Moroi:2002rd}. A variation of the mean value of the axion field across the observable universe is then interpreted as a discrete superhorizon mode \cite{Erickcek:2008sm}.

This paper is organised as follows. In \S~\ref{sec:review} we provide a  derivation of the effect of superhorizon modes on galaxy number counts, complementary to that in Ref.~\cite{Kasai:1987ap}. In \S~\ref{sec:superhorizon} we study the number count fluctuations in the presence of adiabatic and cold dark matter (CDM) isocurvature modes, and also their effects on CMB temperature fluctuations. In \S~\ref{sec:dipoletension} we apply such considerations to the current dipole tension. In \S~\ref{sec:subhorizonmodes} we consider the effect on the number count dipole of modes which are sub-horizon today but were superhorizon at the time of decoupling. We conclude and discuss further directions in \S~\ref{sec:conclusions}. Details of our calculations are provided in the Appendices.
We work throughout in reduced Planck units ($8\pi G=c=1$) unless otherwise stated.

\section{Galaxy number count: review and complementary derivation \label{sec:review}}

We start by computing the general relativistic galaxy number count fluctuations following the covariant formulation of Ellis \cite{Ellis:1971pg} and Kasai \& Sasaki \cite{Kasai:1987ap}.
In Appendix~\ref{app:comparison} we compare our formulation with other recent works \cite{Bonvin:2011bg,Challinor:2011bk,Nadolny:2021hti} and show that they are exactly equivalent. An improvement we make is in not dropping any monopole or dipole terms at the Observer's position.

\begin{center}
\begin{figure}
\includegraphics[width=0.5\columnwidth]{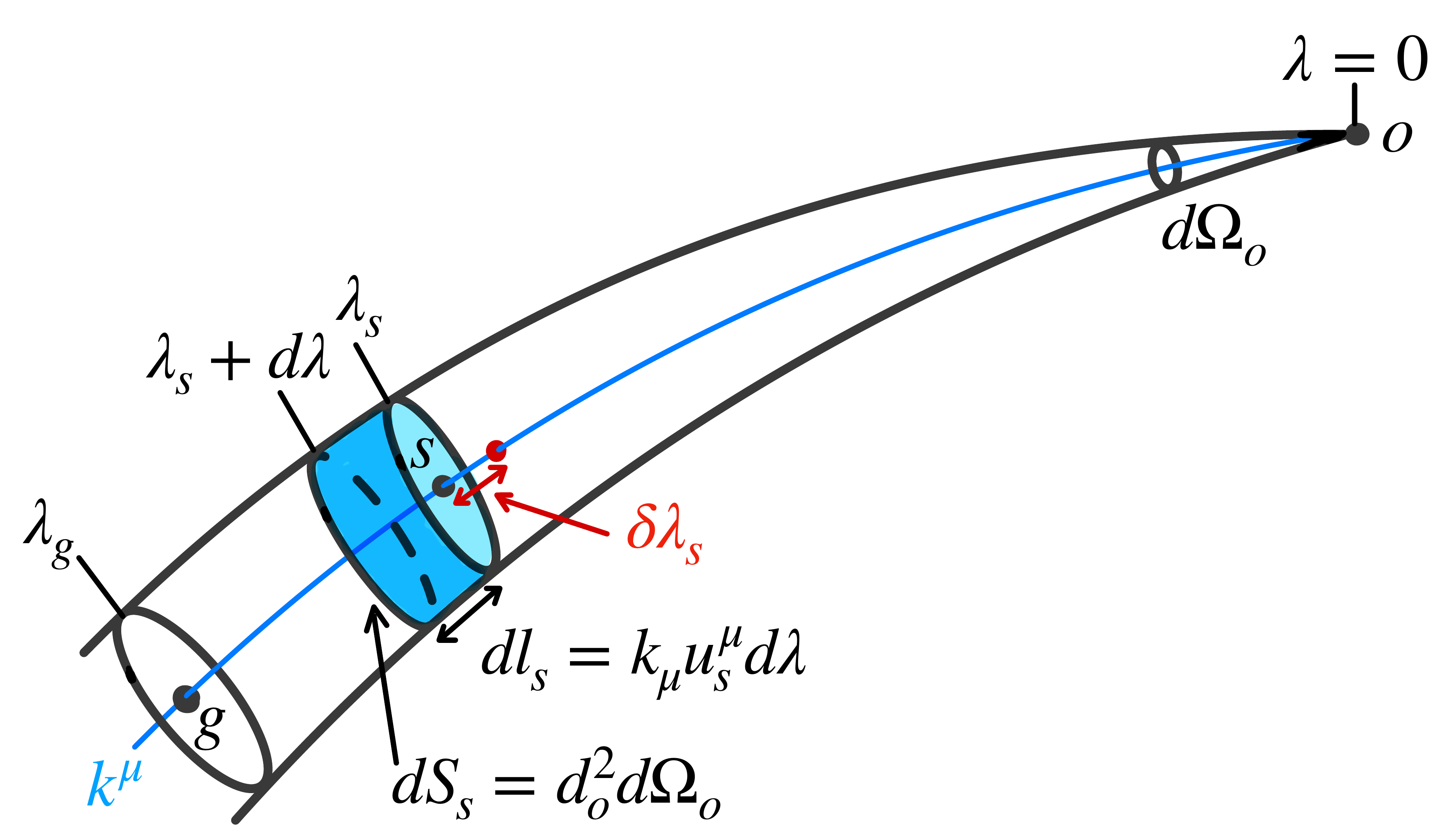}
\caption{A source at $s$ emits photons which reach us at $o$ along null geodesics $k^\mu$ parameterised by the affine parameter $\lambda$. Due to the source velocity $u^\mu_s$, the photon has a spatial displacement $dl_s$  following an infinitesimal change in the affine parameter $d\lambda$. The observed infinitesimal volume is  $dV_s=dl_s dS_s$ and the cross-section $dS_s$ of the photon bundle defines the Observer's area distance $d_o$. 
A fixed affine parameter $\lambda_s$ hypersurface is related to a fixed redshift $z$ hypersurface by shifting the affine parameter of the source by $\delta\lambda_s$. There may also be another source $g$ beyond $s$. The total number count $N$ per solid angle at $o$ is the integral along the whole photon bundle. (After Figure 8 in Ellis \cite{Ellis:1971pg}.)
\label{fig:image}}
\end{figure}
\end{center}

Consider sources of number density $\hat n_s$, with given 4-velocity $\hat u_s^\mu$ at a given point $s$ in a perturbed flat FLRW universe with metric $\hat g_{\mu\nu}$. For the moment, we assume that $\hat n_s$ is a function of time only. The sources emit photons with 4-momentum $\hat k^\mu$, which travel along null geodesics parametrised by an affine parameter $\hat\lambda$ and reach our location at a point $o$ from a direction $\hat n$. We are moving with 4-velocity $\hat u_o^\mu$ so the number of sources we see at point $o$ within the distance $d\hat l_s$, travelled by a photon after an infinitesimal change in the affine parameter $d\hat\lambda$, is \cite{Ellis:1971pg}
\begin{align}
dN (\hat \lambda,\hat n) = \hat n_s\, d\hat l_s d\hat S_s\,,
\end{align}
where $d\hat S_s$ is the cross-sectional area of the photon bundle at $s$. The cross-sectional area $d\hat S_s$ observed at $o$ is independent of the Observer's velocity $\hat u_o^\mu$ if the latter is $\ll c$ \cite{Ellis:1971pg}. The spatial displacement of the photon after $d\hat\lambda$ is $d\hat l_s = (-\hat k_\mu \hat u_s^\mu) d\hat\lambda$, where $\hat \omega = -\hat k_\mu \hat u_s^\mu$ is the energy of the photon. Then the change of the cross-sectional area along the null geodesics is \cite{Ellis:1971pg}
\begin{align}
\frac{d}{d\hat\lambda}(d\hat S_s) = d\hat S_s \hat\nabla_\mu  \hat k^\mu\,.
\end{align}
The cross-sectional area is used to define the Observer's area distance $\hat d_o$ by $d\hat S_s = {\hat d}_o^2 d\Omega_o$ where $d\Omega_o$ is the solid angle of the photon bundle at the Observer's position. Before showing the equations for $\hat k^\mu$ and $\hat d_o$, it is more convenient to 
to work with conformally rescaled variables given by
\begin{align}
\label{eq:conformalvariables}
\hat g_{\mu\nu} = a^{2} g_{\mu\nu} \quad&,\quad 
d\hat\lambda = -a^{2}d\lambda \quad,\quad 
\hat k^\mu = -a^{-2} k^\mu\\ 
\hat u^\mu = a^{-1} u^\mu \quad,\quad 
\hat d_o& = a\, d_o\quad{,}\quad 
\hat n_s = a^{-3} n_s\quad{\rm and}\quad 
\hat\omega = a^{-1}\omega = a^{-1}k_\mu u_s^\mu\,.
\end{align}
In these variables, we have:
\begin{align}
\label{eq:dN}
dN (\lambda,\hat n) = n_s \omega d_o^2d\lambda d\Omega_o\,,
\end{align}
with all the arguments above evaluated at a point parameterised by $(\lambda,\hat n)$. The equations necesssary to solve the system are the null geodesics equation, the Raychaudhuri equation for null geodesics, and the change of cross sectional area, respectively \cite{Kasai:1987ap}
\begin{align}
\frac{dk^\alpha}{d\lambda}& = -\Gamma^\alpha_{\mu\nu}k^\mu k^\nu\,,\label{eq:nullgeodesics}\\
\frac{dd_o}{d\lambda}& = \frac{1}{2}\theta d_o\,,\label{eq:areachange}\\
\frac{d\theta}{d\lambda}& = -R_{\mu\nu}k^\mu k^\nu-\frac{1}{2}\theta^2-\sigma^{\mu\nu}\sigma_{\mu\nu}
\label{eq:geodesicdeviation}\,,
\end{align}
where $\theta\equiv\nabla_\mu k^\mu$ and $\sigma_{\mu\nu}$ are respectively the expansion scalar and the shear of the congruence of null geodesics \cite{Poisson:2009pwt}. Note that due to the conformal invariance of null geodesics, the equations for the original variables would look exactly the same as Eqs.~\eqref{eq:nullgeodesics}-\eqref{eq:geodesicdeviation} but all with ``hatted'' quantities. Also we use the subscripts $s$ and $o$ to respectively denote evaluation at the Source and Observer positions, except for $d_o$ which denotes the area distance to the Source measured by the Observer. This means that for $\lambda$, the affine parameter of the past directed null geodesics, we have $\lambda_o = 0$. Alternatively, $\lambda=0$ means that the Observer is on top of the Source.

\subsection{Number count fluctuations due to cosmological perturbations \label{subsec:perturbations}}

We proceed to derive the galaxy number count fluctuations under the effects of cosmological perturbations. We present the main formulae here and provide the details in Appendix~\ref{app:perturbativenumber}. We perturb the variables as follows. First, following Kasai \& Sasaki \cite{Kasai:1987ap}, we denote all perturbed variables with a tilde, \textit{e.g.}, $\tilde g_{\mu\nu}$ is the perturbed metric and $g_{\mu\nu}$ is the background one. For simplicity, we neglect any vector or tensor perturbation. We work in a perturbed flat FLRW metric in conformal coordinates $(\eta, x^i)$ and in the shear free gauge, \textit{viz.}
\begin{align}
\tilde g_{\mu\nu} = \eta_{\mu\nu} + \delta g_{\mu\nu}, \quad{\rm where}\quad 
\delta g_{\mu\nu} = -2\Psi\delta^0_{(\mu}\delta^0_{\nu)} + 2\Phi\delta_{ij}\delta^i_{(\mu}\delta^j_{\nu)}\,.
\end{align}
The 4-velocities of the Source and the Observer read respectively:
\begin{align}
\tilde u_{s/o}^\mu = (1 - \Psi_{s/o})\delta^\mu_0+ v_{s/o}^i\delta^\mu_i\,.
\end{align}
Similarly, we have:
\begin{align}
\label{eq:deltavariables}
\tilde k^\mu = k^\mu+\delta k^\mu\quad&,\quad \tilde\omega = 1 + \delta\omega\,,\\ 
\tilde d_o = d_o (1 + \delta_r)\quad,\quad 
\tilde n_s = n_s &(1 + \delta_n)\quad{\rm and}\quad 
\tilde\eta = \eta+\delta\eta\,.
\end{align}
The form of $\delta k^\mu$, $\delta_r$, $\delta\omega$ and $\delta\eta$ in terms of $\Phi$, $\Psi$ and $v_{s/o}^i$ are given by the solutions to Eqs.~\eqref{eq:nullgeodesics}-\eqref{eq:geodesicdeviation}. Note that without loss of generality we have normalised the constant energy of the photon to unity, \textit{i.e.} $\omega=1$, since the null geodesics do not depend on the photon energy and we are only interested in relative changes in number counts. We present the solutions below and refer the reader to Appendix~\ref{app:perturbativenumber} for more details. 

Before proceeding further, it is important to note that up to this point we worked in terms of the affine parameter $\lambda$. However, observations are made at a fixed redshift $z = \hat \omega_s/\hat \omega_o = (a_o\omega_s)/(a_s\omega_o)$ and not at a fixed $\lambda$. This means that we base our observations on a measured $\tilde z$ and then associate a ``background'' redshift $z = a_o/a_s$ to it. We take this into account by introducing a shift $\delta\lambda_s$ to the affine parameter $\lambda_s$ of the Source due to cosmological perturbations, which is defined by
\begin{align}
\label{eq:definitionlambdas}
1 + \tilde z(\lambda_s + \delta\lambda_s) \equiv 1 + z(\lambda_s)\,.
\end{align}
With the perturbation expansions \eqref{eq:deltavariables} \& \eqref{eq:definitionlambdas} we find that the total number of sources up to redshift $z$ per unit solid angle is given by
\begin{align}
\label{eq:Nz}
\frac{d\tilde N(z,\hat n)}{d\Omega_o}& = \int_0^{\lambda_s+\delta\lambda_s} d\lambda\,\tilde n_s \tilde\omega_s \tilde d_o^2\nonumber\\& = \int_0^{\lambda_s} d\lambda\, n_s d_o^2+n_s d_o^2(\lambda_s)\delta\lambda_s + \int_0^{\lambda_s}d\lambda \,n_sd_o^2\left(\delta_n + 2\delta_r + \delta\omega_s\right)\,.
\end{align}
Note that in Eq.~\eqref{eq:Nz}, $\lambda_s$ is as in the background relation $z(\lambda_s)$. We use this notation henceforth.

While Eq.~\eqref{eq:Nz} is what current observations measure, we expect that redshift tomography would be possible in future surveys. In anticipation, we compute the number count fluctuations of galaxies per redshift bin, and per solid angle \textit{viz.}
\begin{align}
\label{eq:Deltan}
\Delta_n(z_s, \hat n) \equiv \left(\frac{d N(z_s,\hat n)}{dz_sd\Omega_o}\right)^{-1} \left({\frac{d\tilde N(z_s, \hat n)}{dz_s d\Omega_o} - \frac{d N(z_s,\hat n)}{dz_sd\Omega_o}}\right)\,.
\end{align}
Eq.~\eqref{eq:Deltan} is also the quantity considered in Refs.~\cite{Bonvin:2011bg,Challinor:2011bk}. A straightforward expansion of the derivative of Eq.~\eqref{eq:Nz} with respect to $\lambda_s$ yields
\begin{align}
\label{eq:deltaNfirststep}
\Delta_n(z_s,\hat n) = 2\frac{d\ln d_o}{d\lambda_s}\delta\lambda_s + \frac{d\delta\lambda_s}{d\lambda_s} + \delta_{n,s}+2\delta_{r,s} + \delta\omega_s\,,
\end{align}
where, for the moment, we consider $n_s$ to be constant. 

We present below the solutions to Eqs~\eqref{eq:nullgeodesics}-\eqref{eq:geodesicdeviation} at the background level and at first-order in perturbation theory. The details are presented in Appendix~\ref{app:perturbativenumber}. First, we find at the background level, and along the line of sight, that
\begin{align}
k^\mu = (-1,n^i)\quad,\quad{\eta=\eta_o-\lambda}\quad{\rm and}\quad d_o(\lambda) = r(\lambda) = \lambda\,,
\end{align}
where $n^i$ is the direction of observation of the photon in the comoving radial direction $r$ from the Observer. Then, the first-order solutions read
\begin{align}
\label{eq:deltaomega}
\delta\omega_s &=\Psi_o - \Psi_s + n_iv_s^i-n_iv_o^i - \int_0^{\lambda_s} d\lambda_1\frac{\partial}{\partial \eta}(\Psi - \Phi)\,,\\
\label{eq:deltaeta}
\delta\eta_s &= \lambda_s(n_iv^i_o - \Psi_o) + 2\int_{0}^{\lambda_s}d\lambda_1\Psi + \int_0^{\lambda_s}d\lambda_1\int_0^{\lambda_1}d\lambda_2\frac{\partial}{\partial \eta}(\Psi - \Phi)\,,\\
\label{eq:deltalambda}
\delta\lambda_s &= \delta\eta_s-\frac{\delta\omega_s}{{\cal H}_s}\,,\\
\label{eq:deltar}
\delta_{r,s} &= \Phi_o+\int_0^{\lambda_s}\frac{d\lambda_1}{r^2(\lambda_1)}\int_0^{\lambda_1}d\lambda_2 \,r^2(\lambda_2) \nonumber\\ &\hspace{2cm} \times\left\{\frac{\partial^2}{\partial\eta^2}\Phi - 2\frac{\partial}{\partial\eta}\frac{\partial}{\partial r}\Phi - \frac{1}{2}\Delta(\Psi - \Phi) + \frac{1}{2}\frac{\partial^2}{\partial r^2}(\Psi + \Phi)\right\}\,,
\end{align}
where $\Delta\equiv \partial_i\partial^i$ is the spatial Laplacian. We have set the conditions at the Observer's position to be $\delta\omega_o = \delta\eta_o = \delta_{r,o} = 0$. Any constant $\delta\eta_o$ and $\delta_{r,o}$ can be respectively reabsorbed by a constant shift in the time coordinate and in the affine parameter. The condition $\delta\omega_o = 0$ means that there is no change in the observed photon energy if we are on top of the Source. Eq.~\eqref{eq:Nz}, together with Eqs.~\eqref{eq:deltaomega}-\eqref{eq:deltar} complete our derivation of the number count fluctuations.

\subsection{Corrections due to flux threshold and evolution of sources \label{subsec:corrections}}

So far we have derived the galaxy number count in a perturbed flat FLRW universe assuming that the galaxy comoving number density is constant, and that galaxies have a monochromatic spectrum. However, in actual surveys we usually count the number of galaxies above some flux threshold $F_*$:
\begin{align}
\label{eq:calNs}
\hat {\cal N}_s(z,F>F_*) = \int_{ L_s(z,F)}^\infty\, dL \, \hat n_s(z, \ln L)\,,
\end{align}
where $L_s$ is the source luminosity given by
\begin{align}
\label{eq:L}
L_s=4\pi F \hat d_o^2 (1 + z)^4=4\pi F \tilde d_o^2 a^2(\tilde\eta)(1 + \tilde z)^4\,.
\end{align}
To find the effect of cosmological perturbations, we define $d{\cal N}$ as $dN$ in Eq.~\eqref{eq:Nz} and replace $n_s$ by ${\cal N}_s$. Proceeding as before we obtain
\begin{align}
\label{eq:barNz}
\frac{d {\cal N}(z, \hat n)}{d\Omega_o} &=\int_0^{\lambda_s + \delta\lambda_s} d\lambda\,\tilde{\cal N}_s(\tilde\eta,\tilde L) \tilde\omega \tilde d_o^2\nonumber\\ 
&= \int_0^{\lambda_s} d\lambda\,{\cal N}_s(\eta, L) r^2 + {\cal N}_s(\eta, L)r^2(\lambda_s)\delta\lambda_s\nonumber\\&\hspace{1cm} + \int_0^{\lambda_s}d\lambda\,{\cal N}_s r^2\left(\delta_n + 2\delta_r + \delta\omega + \frac{\partial \ln {\cal N}_s}{\partial \eta}\delta\eta + \frac{\partial \ln {\cal N}_s}{\partial \ln L}\delta \ln L\right)\,,
\end{align}
where we treated $\eta$ and $\ln L$ as separate variables. From Eq.~\eqref{eq:L} we have that $\delta \ln L = 2\delta_r + 2{\cal H}\delta\eta_s - 4{\cal H}\delta\lambda_s$. Taking the derivative of Eq.~\eqref{eq:barNz} with respect to the redshift along the null geodesics, we arrive at
\begin{align}
\label{eq:predeltaN}
\Delta_{{\cal N}}(z_s,\hat n) = &\left(\frac{\partial \ln {\cal N}_s}{\partial \ln (1+z_s)} - 2\frac{\partial \ln {\cal N}_s}{\partial\ln L}\right){\cal H}_s\left(\delta\lambda_s - \delta\eta_s\right)\nonumber\\ 
&+ 2\frac{\partial \ln {\cal N}_s}{\partial\ln L}\left(\frac{d\ln r}{d\lambda_s}\delta\lambda_s + \delta_r\right) + 2\frac{d\ln r}{d\lambda_s}\delta\lambda_s + \frac{d\delta\lambda_s}{d\lambda_s} + \delta_n + 2\delta_{r, s} + \delta\omega_s \,.
\end{align}

To compare with observations we define, as in previous work, the magnitude distribution index and `evolution bias', respectively:
\begin{align}
\label{eq:sandfevo}
x \equiv -\frac{\partial\ln {\cal N}_s}{\partial \ln L}\quad {\rm and}\quad
f_{\rm evo}\equiv - \frac{\partial \ln {\cal N}_s}{\partial \ln (1+z_s)}\,.
\end{align}
For practical purposes, it is more convenient to work with the number density fluctuations in the comoving matter slicing $\delta_{n{\rm c}}$, which are related to the number density fluctuations in the shear free slicing $\delta_{n}$ as \cite{Durrer:2020fza,Nadolny:2021hti}:
\begin{align}
\label{eq:deltanc}
\delta_{n} = \delta_{n{\rm c}} + \frac{\partial \ln \hat {\cal N}_s}{\partial\eta} v_s = b(z)\delta_{m{\rm c}} + (3 - f_{\rm evo}){\cal H}_s v_s\,,
\end{align}
where we used that $v_s^i = \partial^iv_s$.
In Eq.~\eqref{eq:deltanc} we have introduced the linear bias factor $b(z)$, making the usual assumption that the comoving galaxy number count fluctuations $\delta_{n{\rm c}}$ are proportional to the fluctuations  $\delta_{m{\rm c}}$ in the CDM. The advantage of working with the former rather than the latter is that $\delta_{n{\rm c}}$ vanishes for superhorizon adiabatic fluctuations, making it convenient to set initial conditions in the early universe. With Eqs.~\eqref{eq:sandfevo} and \eqref{eq:deltanc}, we rewrite Eq.~\eqref{eq:predeltaN} as
\begin{align}
\label{eq:deltaNready}
\Delta_{{\cal N}}(z_s,\hat n) = &\Delta^\delta_n - 2x(z)\Delta^x_n - f_{\rm evo}(z)\Delta^{\rm evo}_n\,,
\end{align}
where we defined
\begin{align}
\label{eq:deltaN1}
&\Delta^{\delta}_n = b(z)\delta_{mc} + 2\frac{\delta\lambda_s}{\lambda_s} + 2\delta_{r, s} + \frac{d\delta\lambda_s}{d\lambda_s} + 3{\cal H}_s v_s + \delta\omega_s\,,\\
&\Delta_n^x = \delta\omega_s + \frac{\delta\lambda_s}{\lambda_s} + \delta_{r,s}\,,
\label{eq:deltaNs}\\
&\Delta_n^{\rm evo} = {\cal H}_sv_s - \delta\omega_s\,.
\label{eq:deltaNevo}
\end{align}
The purpose of the separation in  Eq.~\eqref{eq:deltaNready} is that in the next section we compute the three terms $\Delta^{\delta}_n$, $\Delta_n^x$ and $\Delta_n^{\rm evo}$, independently of $b(z)$, $x(z)$ and $f_{\rm evo}(z)$, which are harder to determine directly.

\subsection{Kinematic dipole \label{subsec:kinematic}}

First we consider the dipole due to our local motion. The kinematic dipole due to the velocity $v_o$ in $\Delta_{\cal N}$ is given by
\begin{align}
d^{\rm kin}_{n} = \left(2+\frac{{\cal H}'_s}{{\cal H}_s^2} + \frac{2 - 5s}{r_s{\cal H}_s} - f_{\rm evo}\right)\beta\,,
\end{align}
where $\beta\equiv n_iv^i_{o}$. The dipole in total number count ${\cal N}$ is then
\begin{align}
d^{\rm kin}_{\cal N} &= \frac{1}{\int_0^{\lambda_s} {\cal N}_s(\lambda, L) r^2d\lambda}\int_0^{\lambda_s} D^{\rm kin}_{n} {\cal N}_s(\lambda,L) r^2d\lambda\\&
= \frac{1}{\int_0^{\lambda_s} {\cal N}_s(\lambda,L) r^2d\lambda}\int_0^{\lambda_s} \left(2+\frac{{\cal H}'}{{\cal H}^2}+\frac{2(1-x)}{r{\cal H}}+\frac{\partial \ln {\cal N}_s}{\partial \ln (1+z)}\right)\beta {\cal N}_s(\lambda,L) r^2d\lambda\,.
\end{align}
As explained in Appendix~B of Ref.~\cite{Nadolny:2021hti}, we can perform the  integral above by looking for total derivatives, for instance we can use that
\begin{align}
\frac{\partial \ln {\cal N}_s}{\partial \ln (1+z)}&=\frac{d \ln {\cal N}_s}{d \ln (1+z)}-\frac{\partial \ln {\cal N}_s}{\partial \ln L}\frac{\partial \ln L}{\partial \ln (1+z)}=\frac{1}{{\cal H}}\frac{d\ln{\cal N}}{d\lambda}+x\left({1+\alpha}+\frac{2}{r{\cal H}}\right)\,,
\end{align}
where we have made the usual assumption that the sources have a power-law spectrum: $L(\nu)\propto \nu^{-\alpha}$. After integration, we obtain
\begin{align}
d^{\rm kin}_{\cal N} &= \frac{1}{\int_0^{\lambda_s} {\cal N}_s(\lambda,L) r^2d\lambda} \int_0^{\lambda_s} \left(2 + x(1 + \alpha) + \frac{1}{{\cal H}}\frac{d}{d \lambda}\ln\left(\frac{r^2{\cal N}_s}{{\cal H}}\right)\right)\beta {\cal N}_s(\lambda,L) r^2d\lambda
\\ &\qquad\qquad = \left(2 + x(1 + \alpha)\right)\beta + \frac{1}{\int_0^{\lambda_s} {\cal N}_s(\lambda, L) r^2d\lambda}\frac{r_s^2{\cal N}_s}{{\cal H}_s}\beta
\label{eq:Ellisformula}\,.
\end{align}
If we focus on the second term of Eq.~\eqref{eq:Ellisformula} and rewrite it in terms of redshift, we find 
\begin{align}
\label{eq:secondterm}
\frac{1}{\int_0^{\lambda_s} {\cal N}_s(\lambda,L) r^2d\lambda}\frac{r_s^2{\cal N}_s}{{\cal H}_s}\beta\simeq \frac{\beta}{2}\frac{{\cal N}_s(z_s, L(z_s))}{\int^{\eta_0}_{\eta_s} {\cal N}_s(\eta, L(\eta)) \left(1 - {\eta}/{\eta_0}\right)^2d(\eta/\eta_0)}\frac{\left(1 - (1 + z_s)^{-1/2}\right)^2}{(1 + z_s)^{1/2}}\,,
\end{align}
where we have approximated ${\eta}/\eta_0\simeq (1 + z)^{-1/2}$, neglecting the effect of $\Lambda$ for analytical simplicity. Hence if ${\cal N}_s(z,L(z))\propto (1+z)^{\gamma}$ with $\gamma<1/2$, this term \eqref{eq:secondterm} quickly becomes negligible. 

We check if this is indeed the case using the actual redshift distribution of the CatWISE quasars \cite{Secrest:2020has} shown in Fig.~\ref{fig:pdfz} and using the formulae for a matter+$\Lambda$ universe given in Appendix~\ref{subsec:background}. We find that the second term in Eq.~\eqref{eq:Ellisformula} is $< 10^{-3}\beta$, hence 
we recover the standard result \cite{ellisbaldwin}:
\begin{align}
\label{eq:ellisformula2}
d^{\rm kin}_{\cal N} &= \left(2 + x(1 + \alpha)\right)\beta\,.
\end{align}
For the CatWISE catalogue of quasars \cite{Secrest:2020has}, $\langle x \rangle \simeq 1.7$ and $\langle \alpha \rangle \simeq 1.26$. So if the number count dipole and the CMB dipole are both solely due to our local motion, the amplitude of the number count dipole should be about 4 times that of the CMB. However, $d^{\rm kin}_{\cal N}$ is actually found to be a factor of over 2 larger than the expected value \cite{Secrest:2020has}. 
In the second step of Eq.~\eqref{eq:Ellisformula} we assumed that $x$ and $\alpha$ are uncorrelated. In general this might not be the case as pointed out by Dalang \& Bonvin \cite{Dalang:2021ruy}. However this has been directly demonstrated to be inconsequential for the CatWISE quasars \cite{Secrest:2022uvx}
hence we do not consider this possibility further. In the next sections, we investigate whether superhorizon perturbations 
can indeed be the source of the above mismatch between the expected and found dipole.

\begin{figure}
\includegraphics[width=0.6\columnwidth]{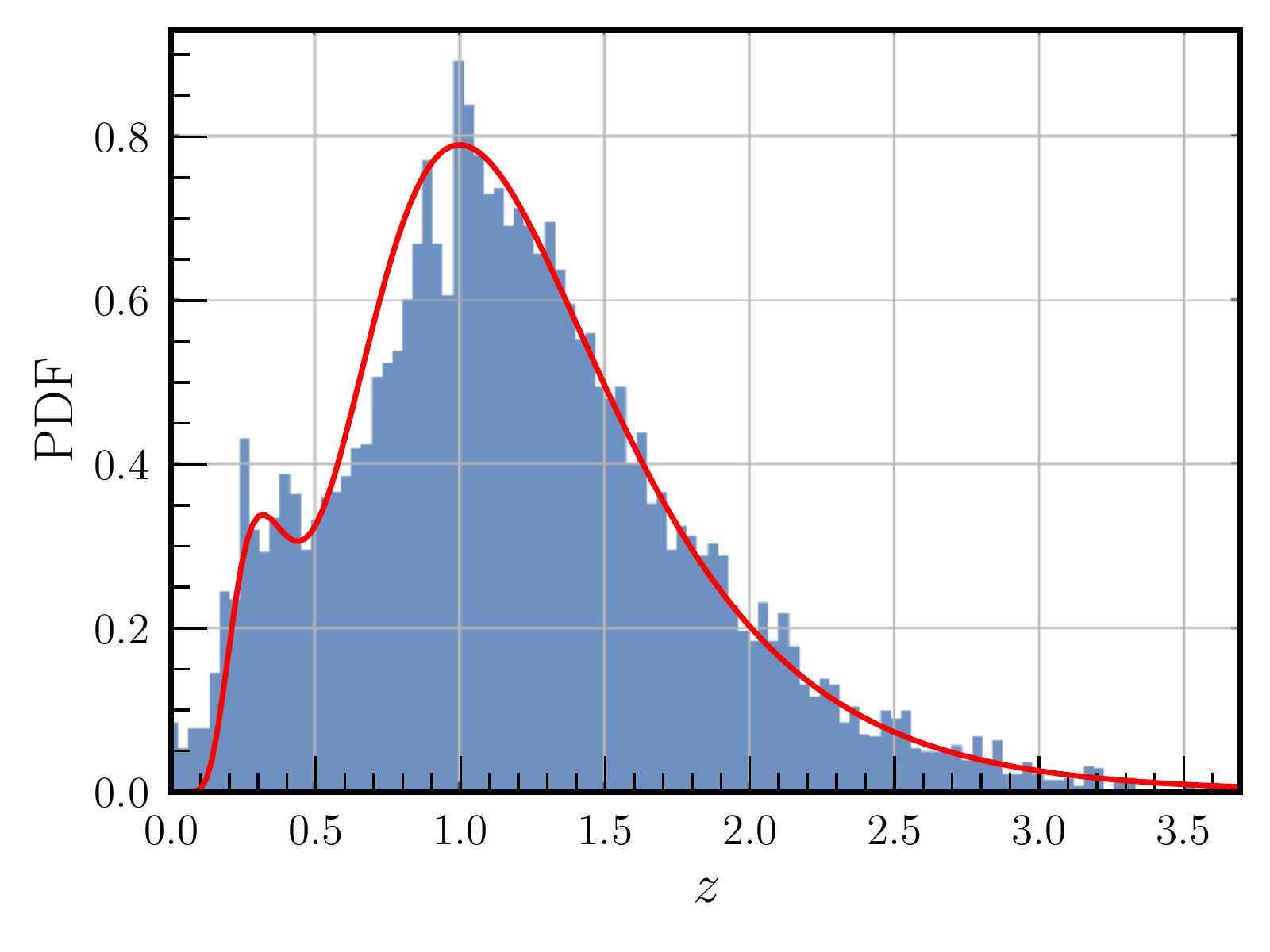}
\caption{Normalised PDF of the CatWISE quasar sample in redshift from Ref.\cite{Secrest:2020has}; 
the comoving number ${\cal N}_s(z)$ is proportional to this PDF times $(1+z)^{-3}$. 
The PDF is fitted with two log-normal distributions as $P(z)=
\left(0.3/\sqrt{2\pi\,\Delta_1^2}\right)
\exp\left(
\log^2[z/0.3]/2\Delta_1^2
\right) + 
\left(0.83/\sqrt{2\pi\,\Delta_2^2}\right) 
\exp\left(-
\log^2[z]/2\Delta_2^2
\right)$, with $\Delta_1=0.37$ and $\Delta_2=0.42$. Since the tail of the distribution falls off exponentially, the term \eqref{eq:secondterm} quickly becomes negligible. 
 \label{fig:pdfz}}
\end{figure}

\section{The effect of superhorizon perturbations today \label{sec:superhorizon}}

We proceed to study the effect of superhorizon modes on the number count fluctuations above some flux limit \eqref{eq:deltaNready}. We also review the effect on the CMB temperature fluctuations as the constraints from CMB observations play a crucial role. However, before going into the details of the calculations, it is helpful to briefly review the possible initial conditions in the early universe and the evolution of superhorizon fluctuations until the present day. For analytical tractablility, we consider the expansion history of the universe in terms of known analytical solutions, \textit{i.e.} we separately treat a radiation + matter universe and a matter + $\Lambda$ universe. The former is important for the CMB fluctuations and the latter for the number count fluctuations. In this section, we neglect baryons for simplicity and use the best-fit Planck values \cite{Planck:2018vyg} for various parameters: $z_{\rm eq}\simeq 3400$, $z_{\rm dec} \simeq 1090$, ${\cal H}_{\rm eq} = k_{\rm eq} \simeq 0.0104\,{\rm Mpc}^{-1}$, ${\cal H}_{\rm dec} = k_{\rm dec} \simeq 4.8 \times 10^{-3}\,{\rm Mpc}^{-1}$, $H_0\simeq 67.4\,{\rm km/s/Mpc} \simeq 2.2\times10^{-4}\,{\rm Mpc}^{-1}$, $\Omega_{{\rm m},0}\simeq 0.315$ and $\Omega_{\Lambda,0} \simeq 0.685$. Also, from now on, we split our velocity (or the Sun's velocity $v_\odot^i$) into a peculiar term and the superhorizon contribution:
\begin{align}\label{eq:vodot}
n_iv^i_{\odot}=n_iv^i_o+n_iv^i_m\,.
\end{align}
In Eq.~\eqref{eq:vodot} $n^iv_o^i$ is the peculiar velocity that affects the kinematic dipole in the CMB and the number count as in \S~\ref{subsec:kinematic} and $v_m^i$ is the velocity induced due to the gradient of the superhorizon mode. 
The velocity $v_{o}^i$ should be understood as our velocity with respect to the CRF.

\subsection{Adiabatic vs isocurvature initial conditions}

In the standard cosmological model, initial conditions are set in the very early universe on scales much larger than the particle horizon. In broad terms, there are two possibilities. The first is the so-called adiabatic initial conditions where there are only curvature fluctuations on slices of homogeneous total energy density. The second one is isocurvature initial conditions where, by definition, curvature fluctuations are initially vanishing and there are only relative number density fluctuations. Since we are interested on scales that are still superhorizon today, we neglect baryons for simplicity so that we can treat the early universe as being filled with just radiation and CDM. (Allowing for baryons will introduce changes of ${\cal O}(10)\%$ which is unimportant for our purposes.) In that case, there is only one type of isocurvature perturbation $S$ defined by
\begin{align}
\label{eq:Sdef}
S\equiv\delta_{\rm m} - \frac{3}{4}\delta_{\rm r}\,,
\end{align}
where $\delta_{\rm m}$ and $\delta_{\rm r}$ respectively are the density contrast of matter and radiation, related to the energy densities by: $\delta_{\rm m} = \delta\rho_{\rm m}/\rho_{\rm m}$ and $\delta_{\rm r} = \delta\rho_{\rm r}/\rho_{\rm r}$. In the shear-free gauge, the 00-component of Einstein's equations for superhorizon modes yields (see Appendix ~\ref{app:einstein}, Eq.~\eqref{eq:00EE})
\begin{align}
\label{eq:00}
2\Phi = \frac{\rho_{\rm m}\delta_{\rm m}+\rho_{\rm r}\delta_{\rm r}}{\rho_{\rm r} + \rho_{\rm m}}\,.
\end{align}
Adiabatic initial conditions are characterised by $S_i=0$ and therefore $\delta_{{\rm m},i} = \frac{3}{4}\delta_{{\rm r},i}$, where the subscript ``$i$'' refers to evaluation at the initial time. In the very early universe when $\rho_{\rm r}\gg\rho_{\rm m}$ Eq.~\eqref{eq:00} implies $\delta_{{\rm r},i} = 2\Phi_i$ or, equivalently, $\delta_{{\rm rc}, i} = \delta_{{\rm r}, i} - 4{\cal H}v_{\rm r} = 0$ and, similarly, $\delta_{{\rm mc}, i}=0$. On the other hand, isocurvature initial conditions are characterised by $\Phi_i = 0$. Eq.~\eqref{eq:00} yields: $\rho_{{\rm m},i}\delta_{{\rm m}, i} + \rho_{{\rm r}, i}\delta_{{\rm r}, i} = 0$, thus in the very early universe we have $\delta_{{\rm m},i} = S_i$.

\begin{figure}
\includegraphics[width=0.49\columnwidth]{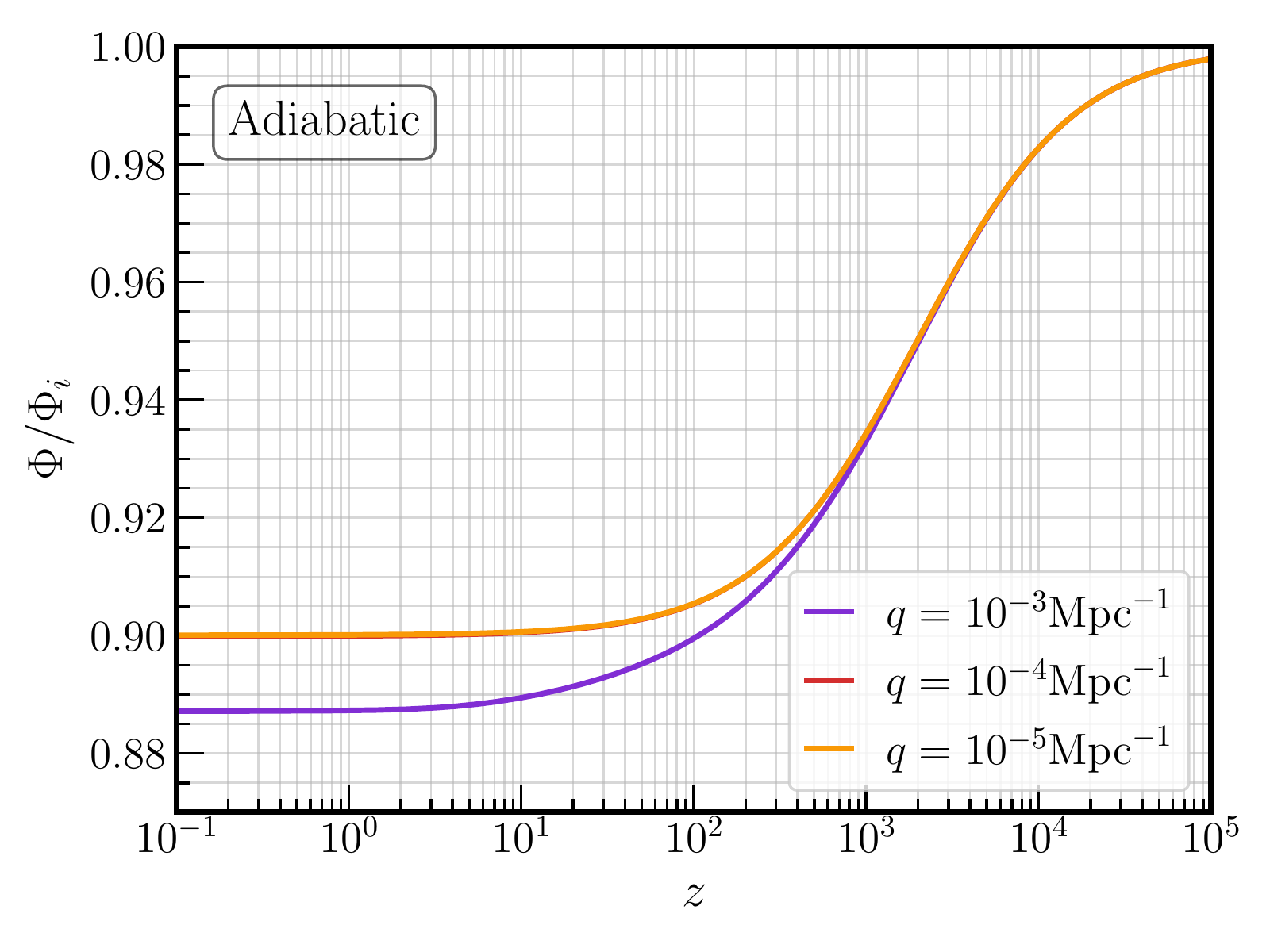}
\includegraphics[width=0.49\columnwidth]{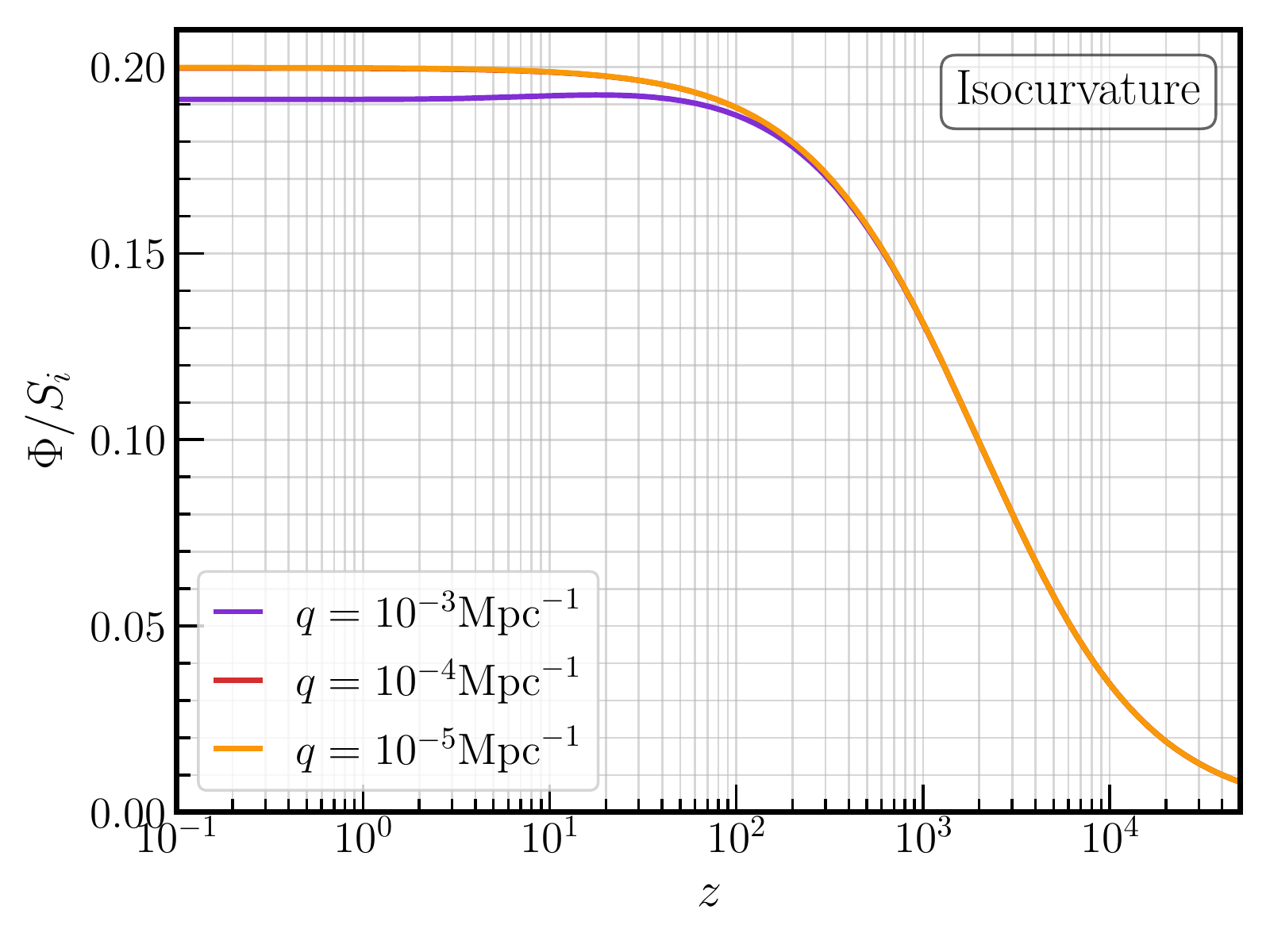}
\includegraphics[width=0.49\columnwidth]{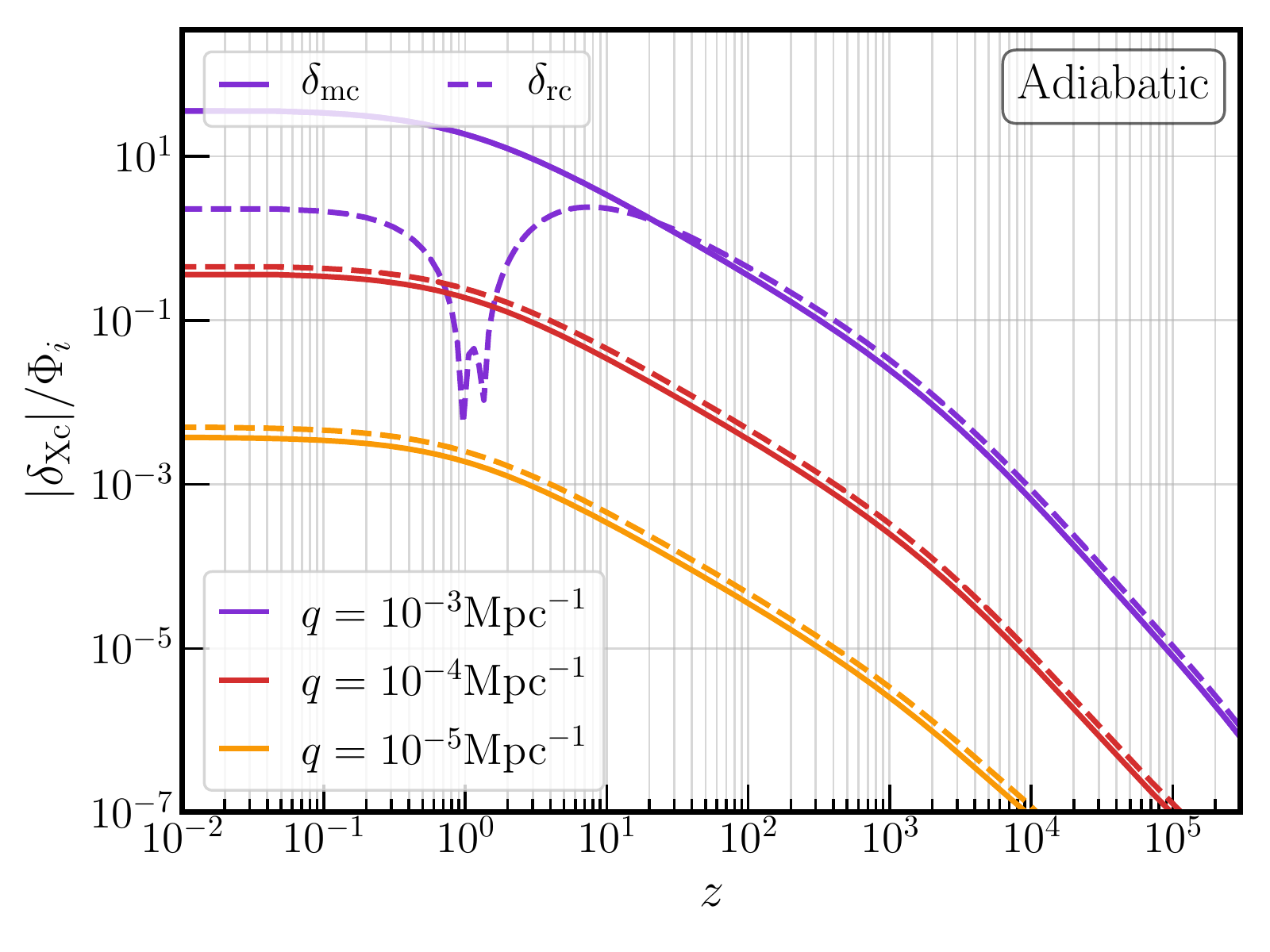}
\includegraphics[width=0.49\columnwidth]{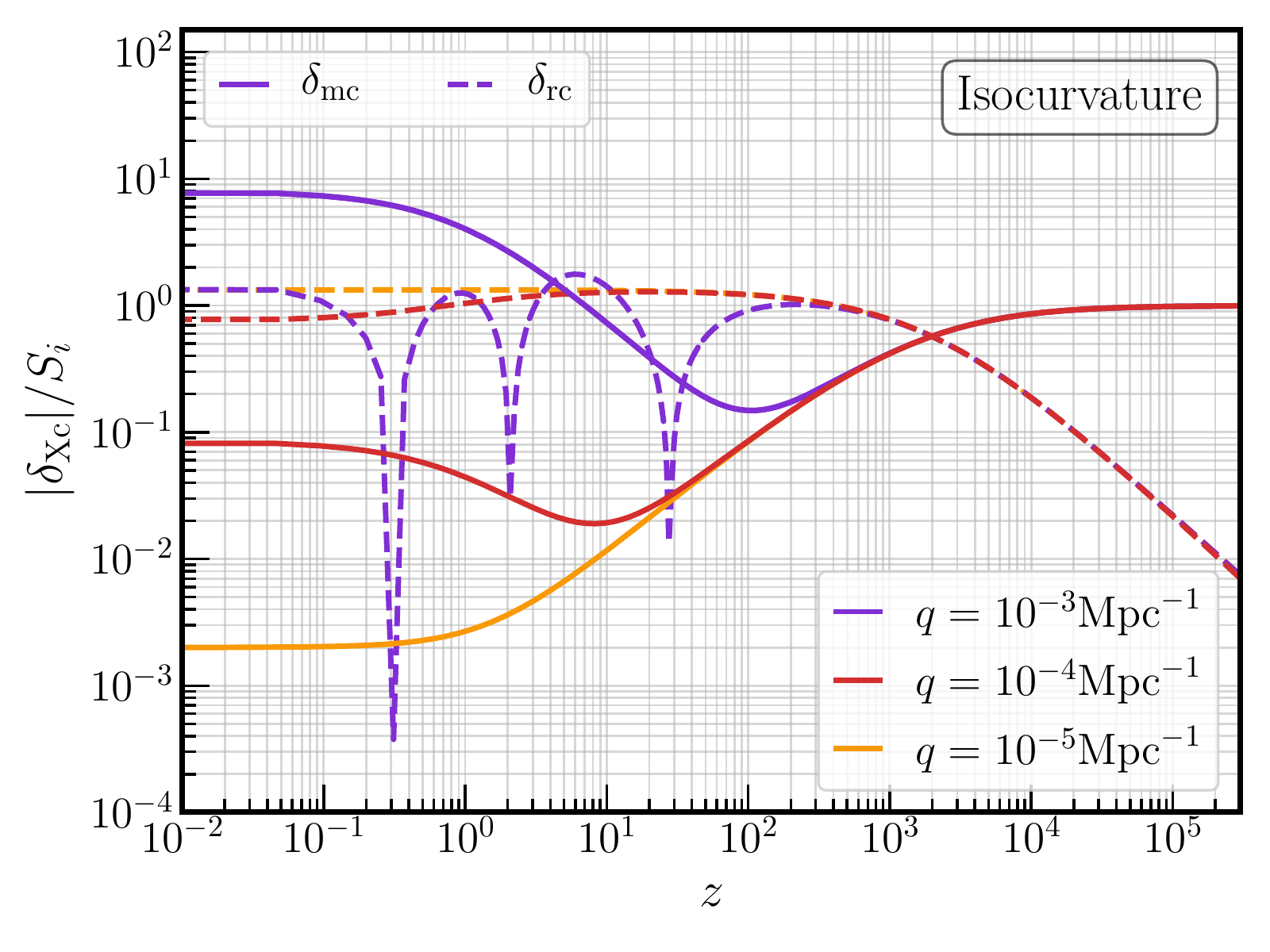}
\caption{Transfer functions for $\Phi$, $\delta_{\rm mc}$ and $\delta_{\rm rc}$ as a function of redshift. The top row shows $\Phi(z)$ for adiabatic (left panel) and isocurvature (right panel) initial conditions --- for $q=10^{-3}\,{\rm Mpc}^{-1}$ (purple),  $q=10^{-4}\,{\rm Mpc}^{-1}$ (red) and  $q=10^{-5}\,{\rm Mpc}^{-1}$ (orange). Note that for superhorizon modes in the isocurvature case, $\Phi=1/5 S_i$ already at redshifts $z\lesssim100$. For slightly subhorizon modes the curvature perturbation has decayed a little for both initial adiabatic and isocuravture perurbations. The bottom row shows the evolution of $\delta_{\rm mc}$ (solid lines) and $\delta_{\rm rc}$ (dashed lines) for adiabatic (left panel) and isocurvature (right panel) initial conditions. 
For superhorizon modes, the density contrast $\delta_{\rm mc}$ of CDM is suppressed at low redshift and the initial isocurvature is transferred to radiation. For slightly subhorizon modes, the density contrast in CDM grows when the mode enters the horizon. For $q= 10^{-3}\,{\rm Mpc}^{-1}$, the density contrast $\delta_{\rm mc}$ is $\sim10$ times the initial curvature $\Phi_i$ or isocurvature $S_i$ perturbation. For $q>10^{-4}\,{\rm Mpc}^{-1}$, the amplitude of the initial curvature and isocurvature perturbations are constrained by the CMB data to be $\Phi_i<10^{-4}$ and $S_i<10^{-5}$ \cite{Planck:2018jri}. \label{fig:plotsphideltaiso}}
\end{figure}

Concerning the galaxy number count fluctuations, we have to evolve the initial fluctuations down to the redshifts of the  surveys which extend to $z \sim 1-2$. This means that we can take the expansion solution for the radiation+matter universe well past the epoch  of radiation-matter equality, \textit{i.e.} $a/a_{\rm eq} \gg 1$. In this limit, and for superhorizon fluctuations, we find (see Appendix~\ref{subsec:perturbations}):
\begin{align}
\label{eq:solutionslatetime}
\Phi(a\gg a_{\rm eq}) &\simeq  \frac{9}{10}\Phi_i + \frac{1}{5}S_i\quad,\quad\delta_{{\rm mc}}(a\gg a_{\rm eq}) \simeq 4\frac{a_{\rm eq}}{a}S_i\quad,\quad
\delta_{{\rm rc}}(a\gg a_{\rm eq}) \simeq - \frac{4}{3}S_i\,,\\
v_{\rm m}(a \gg a_{\rm eq})& \simeq \left(\frac{3}{5}\Phi_i + \frac{2}{15}S_i\right)\frac{\sqrt{2}}{{\cal H}_{\rm eq}}\sqrt{\frac{a}{a_{\rm eq}}}\quad{\rm and}\quad
v_{\rm r}(a\gg a_{\rm eq})\simeq\left(\frac{3}{5}\Phi_i + \frac{4}{5}S_i\right)\frac{\sqrt{2}}{{\cal H}_{\rm eq}}\sqrt{\frac{a}{a_{\rm eq}}}\,.
\end{align}
Eq.~\eqref{eq:solutionslatetime} illustrates that initial isocurvature fluctuations have been converted to adiabatic fluctuations. We show the evolution of $\Phi$, $\delta_{\rm mc}$ and $\delta_{\rm rc}$ in Fig.~\ref{fig:plotsphideltaiso}. Now, despite the difference in the relative velocities of matter and radiation, to evaluate the galaxy number count fluctuations we consider an initial curvature fluctuation in the matter domination (MD) era given by
\begin{align}\label{eq:phiMD}
\Phi_{\rm MD}\equiv \frac{9}{10}\Phi_i+\frac{1}{5}S_i\,,
\end{align}
From Eqs.~\eqref{eq:solutionslatetime} and \eqref{eq:phiMD}, it also follows that
\begin{align}\label{eq:vm}
v_{\rm m}&\simeq\Phi_{\rm MD}\frac{2\sqrt{2}}{3{\cal H}_{\rm eq}}\sqrt{\frac{a}{a_{\rm eq}}}\,.
\end{align}
Eqs.~\eqref{eq:phiMD} and \eqref{eq:vm} show  that for all practical purposes the fluctuations in matter are adiabatic. We take these equations as initial conditions for the matter+$\Lambda$ universe. 

At this point, it is important to note the following. The transfer from isocurvature to curvature perturbations has not been fully completed by today so there remains some density fluctuations of CDM from the initial isocurvature perturbation, as can be seen in the top panel of Fig.~\ref{fig:plotsphideltaiso}. However, at redshifts $z \lesssim2$, the density contrast is suppressed by a factor of $a_0/a_{\rm eq}=(1+z_{\rm eq}) \sim 10^3$. Therefore, we can neglect such traces of initial isocurvature in the galaxy number counts as these cannot possibly account for the observed galaxy number count dipole.

On the other hand, we should consider the effect of modes which were superhorizon at the time of decoupling but are subhorizon today, \textit{i.e.} for modes with $10^{-4}\,{\rm Mpc}^{-1}<q<5\times 10^{-3}\,{\rm Mpc}^{-1}$ which we call `slightly subhorizon'. However, while the amplitude of superhorizon modes today is \textit{a priori} poorly  constrained, the amplitude of slightly subhorizon modes is bounded by CMB data to be $\Phi_i<10^{-4}$ and $S_i<10^{-5}$ \cite{Planck:2018jri}. Thus, despite the fact that such subhorizon modes can potentially affect the number count dipole, as their density contrast today has grown with respect to superhorizon modes (see the bottom panel of Fig.~\ref{fig:plotsphideltaiso}), their contribution is not sufficient to explain the number count dipole anomaly. We present more details in \S~\ref{sec:subhorizonmodes} and show that the dipole contribution from a discrete adibatic, slightly subhorizon mode can at most be of ${\cal O}(10^{-3})$.

\subsection{Effect on the number count fluctuations\label{subsec:numbercount}}

We proceed to study the effect of such a discrete adiabatic superhorizon mode on the galaxy number count fluctuations. If it is still superhorizon today, we can expand the curvature perturbation in powers of $\vec{q}\cdot\vec{r}$ as
\begin{align}\label{eq:phiMDexpansion}
\Phi_{\rm MD}(\vec{q}\cdot\vec{x})=\sum_{\ell=1}^\infty \frac{1}{\ell!}\Phi_{{\rm MD},\ell}(qr\cos\theta)^\ell\,,
\end{align}
where $\Phi_{{\rm MD},\ell}=9\Phi_{i,\ell}/10+S_{i,\ell}/5$, $q$ is the comoving wavenumber of the superhorizon mode and $r$ is the comoving distance to the Source and $\theta$ is the angle between the direction of $\vec{q}$ and $\vec{r}$. We then expand the number count fluctuations \eqref{eq:deltaNready} similarly as
\begin{align}
\label{eq:deltaNexpansion}
\Delta^X_n=\sum_{\ell=1}^\infty \frac{1}{\ell!}D^X_\ell(\cos\theta)^\ell\,,
\end{align}
where $X=\{\delta,x,{\rm evo}\}$. We will compute $D^X_\ell$ up to $\ell=3$. Then, we consider that
\begin{align}
\Phi(a)=\Phi_{\rm MD}(\vec{q}\cdot\vec{x}){\cal T}_{\Phi}(a) \quad{\rm and}\quad v_{\rm m}(a)=\Phi_{\rm MD}(\vec{q}\cdot\vec{x}){\cal T}_{v}(a)\,,
\end{align}
where ${\cal T}_{\Phi}(a)$ and ${\cal T}_{v}(a)$ are respectively the transfer functions for $\Phi$ and $v_{\rm m}$ given by (see Eqs.~\eqref{eq:Phia} and \eqref{eq:va})
\begin{align}
{\cal T}_\Phi(a)&=\frac{5\Omega_{\rm m,0}}{2}\frac{{\cal H}/{\cal H}_0}{a^2}\int_0^{a}\frac{d a_1}{({\cal H}(a_1)/{\cal H}_0)^3}\,,\\
{\cal T}_v(a)&=\frac{5\Omega_{\rm m,0}}{2}\frac{1}{a^2{\cal H}_0}\int_0^a d  a_1 \frac{a- a_1}{ a_1}\frac{1}{({\cal H}( a_1)/{\cal H}_0)^3}\,.
\end{align}
In a dust-dominated flat universe \textit{i.e.} for $\Omega_{\rm m,0}=1$, we have 
\begin{align}
\label{eq:dustdominatedT}
{\cal T}^{\rm dust}_\Phi(a)=1\quad{\rm and}\quad {\cal T}^{\rm dust}_v(a)={\eta}/{3}\,.
\end{align}

Let us focus first on the effect of a superhorizon mode dipole, \textit{i.e.} the $\ell=1$ component of Eq.~\eqref{eq:phiMDexpansion}, on the dipole of Eq.~\eqref{eq:deltaNexpansion} $\Delta_{\cal N}$. Note that because of the presence of a Laplacian in $\delta_r$ \eqref{eq:deltaNready}, we also have a contribution to $\ell=1$ in Eq.~\eqref{eq:deltaNexpansion} from the $\ell=3$ mode in Eq.~\eqref{eq:phiMDexpansion}. This however is of order $(qr)^3$, the form of which we present later. For simplicity, we start with the dust-dominated universe with the transfer functions given by Eq.~\eqref{eq:dustdominatedT}. In this case 
\begin{align}\label{eq:deltadust}
\delta\omega_s=\frac{2}{3}q \lambda_s\mu\quad&,\quad\delta\eta_s=q\lambda_s\mu\left(\frac{1}{3}\eta_0-\lambda_s\right)\quad,\quad \delta_{\rm mc}=0\,,\nonumber\\ \delta\lambda_s=-\frac{2}{3}q\lambda_s\mu\quad&,\quad\delta_r=0\quad{\rm and}\quad {\cal H}_sv_s=\frac{1}{3}q\lambda_s\mu\,,
\end{align}
where we defined $\mu\equiv\cos\theta$ and used $r=\lambda=\eta_0-\eta$. Inserting Eq.~\eqref{eq:deltadust} into Eq.~\eqref{eq:deltaNready} it is straightforward to check that $D^{\delta}_{1}=D^{x}_{1}=D^{{\rm evo}}_{1}={\cal O}((qr)^3)$.
This means that, in a dust-dominated universe, there is no dipole in the galaxy number count fluctuations from a dipole in an adiabatic superhorizon mode. 

This result also holds for the matter+$\Lambda$ universe but to  demonstrate this requires more work. First, it is more convenient to use the expressions for $\delta\eta$ \eqref{eq:deltaetaapp} and $\delta_r$ \eqref{eq:deltarapp} in Appendix~\ref{app:comparison}. Then, we insert the $\ell=1$ term of Eq.~\eqref{eq:phiMDexpansion} into Eq.~\eqref{eq:deltaomega}, Eq.~\eqref{eq:deltaetaapp} and Eq.~\eqref{eq:deltarapp}. Using the formulae \eqref{eq:ISWd} and \eqref{eq:ISW2} we can do some of the integrals to arrive at:
\begin{align}
\label{eq:deltaomegaLambda}
\delta\omega_s&=-q\mu\lambda_s {\cal T}_\Phi(a)+\frac{2}{3\Omega_{{\rm m},0}}q\mu\lambda_s\,,\\\label{eq:deltaetaLambda}
\delta\eta_s&=q\mu\lambda_s {\cal T}_v(1)-2q\mu\int_0^{\lambda_s} d\lambda_1(\lambda_s-\lambda_1)q\mu{\cal T}_\Phi(a)\,,\\
\delta_r&=q\mu\lambda_s{\cal T}_\Phi(a)+\frac{q\mu}{{\cal H}_0}\left[\frac{1}{a^2}\int_0^a da_1\frac{a^2-a_1^2}{a_1^2({\cal H}(a_1)/{\cal H}_0)^3}\right]^{a=a_s}_{a=1}+\frac{2}{\lambda_s}\int_0^{\lambda_s}d\lambda_1(\lambda_s-\lambda_1)q\mu{\cal T}_\Phi(a)\,.\label{eq:deltarLambda}
\end{align}
Plugging in Eqs.~\eqref{eq:deltaomegaLambda}-\eqref{eq:deltarLambda} into Eq.~\eqref{eq:deltaNready} and using the integrals \eqref{eq:ISW3} and \eqref{eq:ISW4}, we find
\begin{align}
\label{eq:nodipole}
D^{\delta}_{1}=D^{x}_{1}=D^{{\rm evo}}_{1}={\cal O}((qr)^3)\,.
\end{align}
Hence there is no leading order dipole contribution to the galaxy number count fluctuations from an adiabatic superhorizon mode dipole, even in a general matter+$\Lambda$ universe. 

This is similar to the case of the CMB, where no adiabatic superhorizon mode affects the CMB dipole at leading order \cite{Turner:1991dn}. Since the dipole in the galaxy number counts and in the CMB arise at the same order, it is thus not possible for an adiabatic superhorizon mode to explain the dipole tension. We now explore the possibility of an isocurvature superhorizon mode which can affect the CMB dipole, but not the number count dipole.

\begin{figure}
\includegraphics[width=0.6\columnwidth]{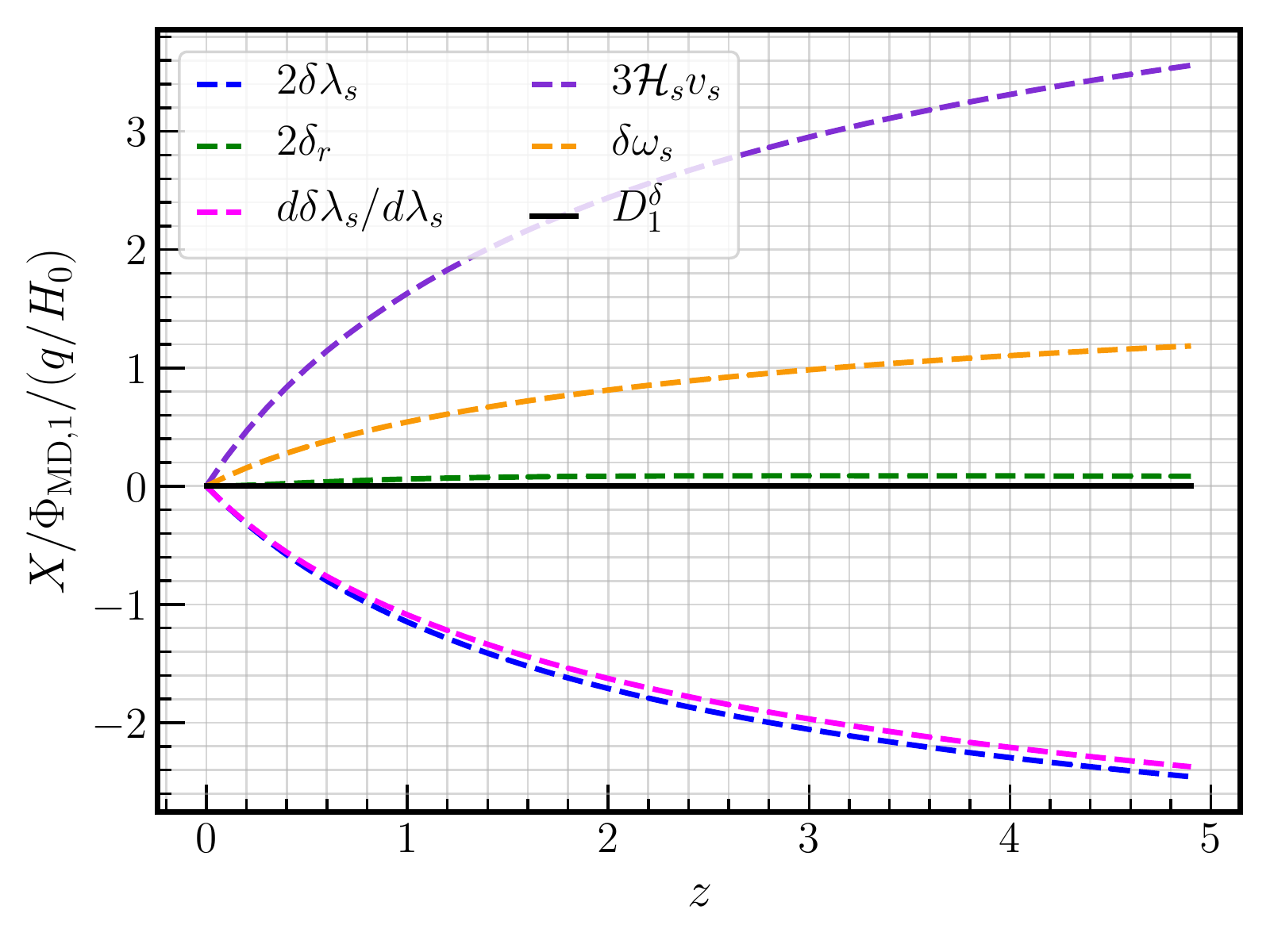}
\caption{Different contributions to the number count dipole $D^\delta_1$ \eqref{eq:deltaNexpansion}. Dashed lines show the different terms in $\Delta^\delta_n$ \eqref{eq:deltaN1}, which are respectively $\delta\lambda_s$ (blue), $\delta_r$ (green), $d\delta\lambda_s/d\lambda_s$ (magenta), ${\cal H}_sv_s$ (purple) and $\delta\omega_s$ (orange), while their sum is shown as the solid black line. Note there is an exact cancellation and $D^\delta_1=0$ at leading order for adiabatic initial conditions. Similar cancellations occur for the other contributions to the dipole, \textit{i.e.} $D^x_1=0$ and $D^{\rm evo}_1=0$. \label{fig:dipolezero}}
\end{figure}

\begin{figure}
\includegraphics[width=0.49\columnwidth]{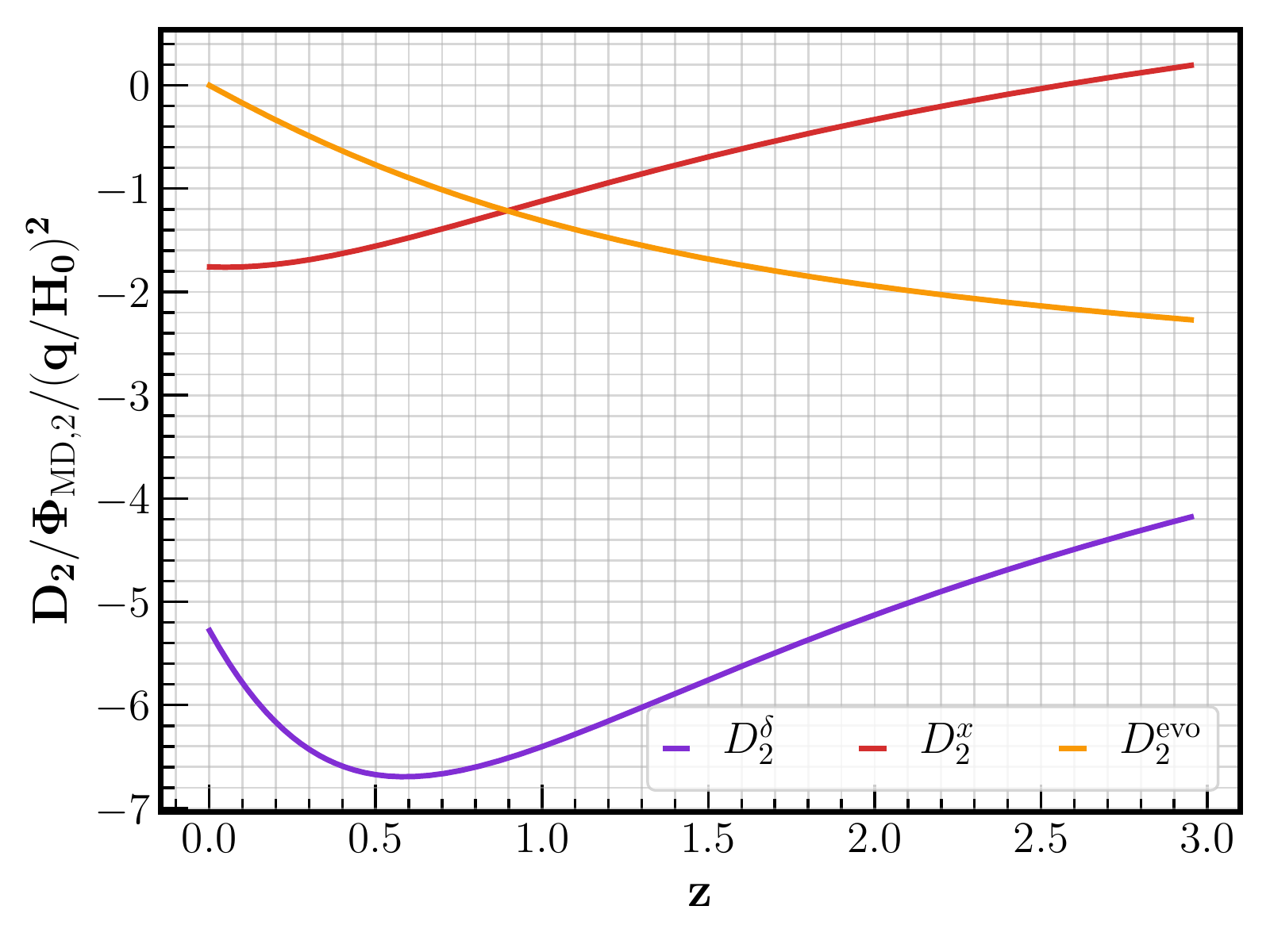}
\includegraphics[width=0.49\columnwidth]{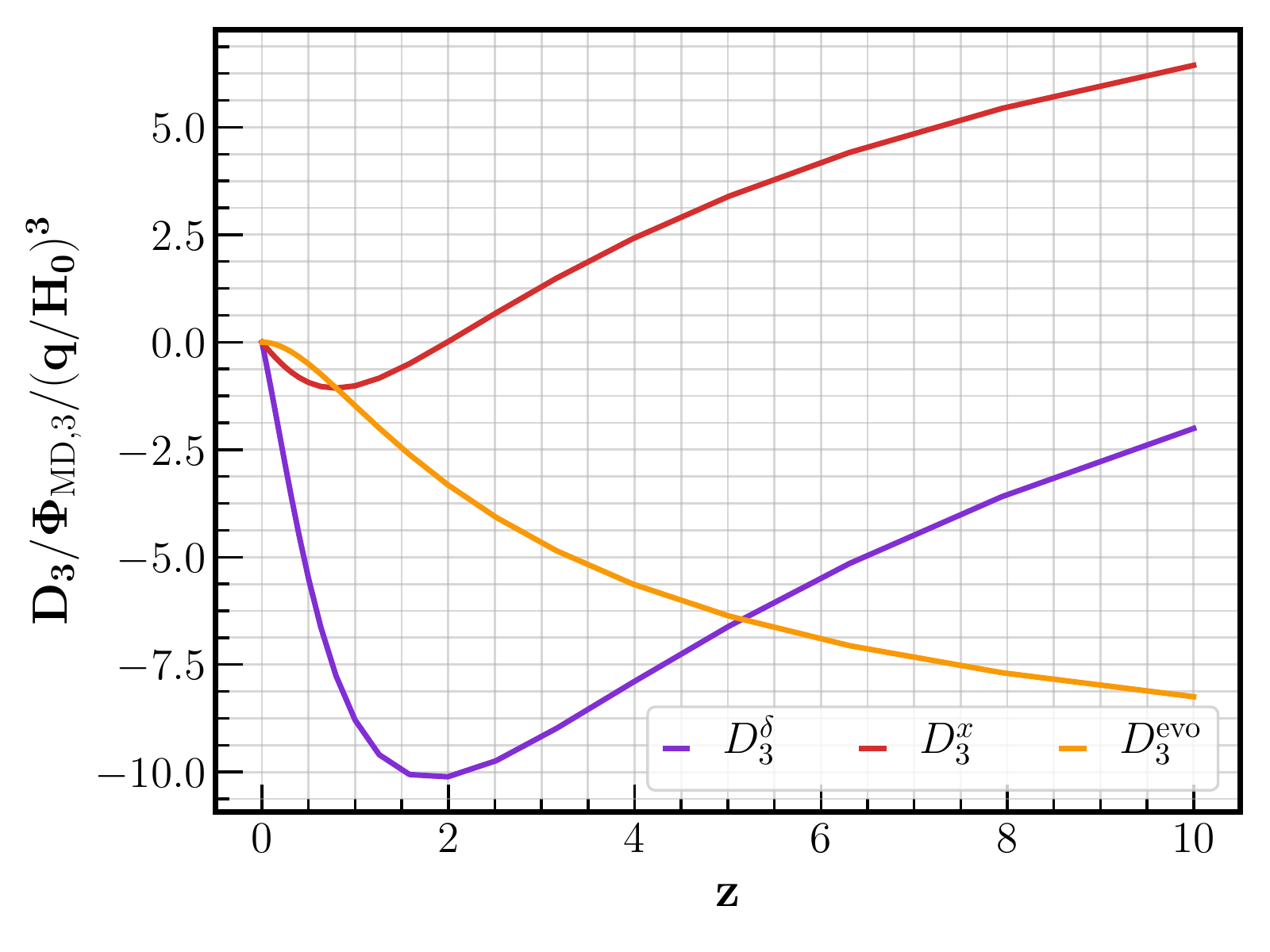}
\caption{Evolution of the galaxy number count quadrupole (left) and octupole (right)  with redshift $z$ in the matter+$\Lambda$ universe. Purple, red and orange lines show respectively the quadrupole and octupole contributions to $\Delta_n^\delta$ \eqref{eq:deltaN1}, $\Delta_n^x$ \eqref{eq:deltaNs} and $\Delta_n^x$ \eqref{eq:deltaNevo}. The fact that the quadrupole does not vanish at $z=0$ implies an anisotropy in the Hubble parameter, see \S~\ref{sec:dipoletension}. \label{fig:quadrupoleoctupole}}
\end{figure}

The contribution from higher $\ell$ terms to the dipole, quadrupole and octupole have to be computed numerically for a general matter+$\Lambda$ universe and are shown in Figs.~\ref{fig:quadrupoleoctupole}. Here we present the analytical results for a matter-dominated universe. First, the contribution to $\ell=1$ is given by
\begin{align}
D^\delta_1=\frac{1}{6}(q\lambda_s)^3\Phi_{{\rm MD},3}\quad,\quad
D^x_1=\frac{1}{12}(q\lambda_s)^3\Phi_{{\rm MD},3}\quad{\rm and}\quad
D^{\rm evo}_1=0\,.
\end{align}
Then, the contribution to the quadrupole reads
\begin{align}\label{eq:quadrupolesincount}
D^\delta_2&=-(q\lambda_s)^2\Phi_{{\rm MD},2}\left(\frac{1}{6}-\frac{8}{3\lambda_s{\cal H}_0}+\frac{4}{\lambda_s^2{\cal H}_0^2}\right)\,,\\
D^x_2&=-(q\lambda_s)^2\Phi_{{\rm MD},2}\left(\frac{1}{6}-\frac{5}{3\lambda_s{\cal H}_0}+\frac{4}{3\lambda_s^2{\cal H}_0^2}\right)\,,\\
D^{\rm evo}_2&=2(q\lambda_s)^2\Phi_{{\rm MD},2}\left(\frac{1}{6}-\frac{2}{3\lambda_s{\cal H}_0}\right)\,.
\end{align}
For the octupole we obtain 
\begin{align}
D^\delta_3&=-(q\lambda_s)^3\Phi_{{\rm MD},3}\left(1-\frac{7}{\lambda_s{\cal H}_0}+\frac{8}{\lambda_s^2{\cal H}_0^2}\right)\,,\\
D^x_3&=-(q\lambda_s)^3\Phi_{{\rm MD},3}\left(\frac{1}{2}-\frac{3}{\lambda_s{\cal H}_0}+\frac{2}{\lambda^2{\cal H}_0^2}\right)\,,\\
D^{\rm evo}_3&=(q\lambda_s)^3\Phi_{{\rm MD},3}\left(\frac{2}{3}-\frac{2}{\lambda_s{\cal H}_0}\right)\,.
\end{align}

Thus, a superhorizon mode predicts a quadrupole in the total number count:
\begin{align}
Q_{\cal N}&=\frac{1}{2}\frac{1}{\int_0^{\lambda_s} {\cal N}_s(\lambda,L) r^2d\lambda}\int_0^{\lambda_s} \left(D^{\delta}_{2}-2xD^{x}_{2}-f_{\rm evo}D^{\rm evo}_{2}\right) {\cal N}_s(\lambda,L) r^2d\lambda\sim -\frac{2}{3}\Phi_{{\rm MD},2}\left(\frac{q}{H_0}\right)^2\,,
\end{align}
where we assumed that most of the contribution  comes from $z<1$ because the quadrupole approaches a constant for $z=0$, and we used that
\begin{align}
-f_{\rm evo}(z\ll1)\simeq x(1+\alpha)+\frac{2(x-1)}{{\lambda {\cal H}}}+1-\frac{3}{2}\Omega_{\rm m,0}+\frac{1}{{\cal H}}\frac{d}{d \lambda}\ln\left(\frac{r^2{\cal N}_s}{{\cal H}}\right)\,.
\end{align}

If we expand Eq.~\eqref{eq:deltaNready} in spherical harmonics instead of an expansion in powers of $(qr_s)$, \textit{viz.}
\begin{align}
\Delta^X_n=\sum_{\ell=1}^\infty \frac{1}{\ell!}D^X_\ell(\cos\theta)^\ell=\sum_{\ell=1}^\infty a^X_{\ell0}Y_{\ell0}(\theta,\varphi)\,,
\end{align}
we have the following translation
\begin{align}
a^X_{10}=2\sqrt{\frac{\pi}{3}}D^X_1+\frac{1}{5}\sqrt{\frac{\pi}{3}}D^X_3\quad,\quad a^X_{20}=\frac{2}{3}\sqrt{\frac{\pi}{5}}D^X_2\quad{\rm and}\quad a^X_{30}=\frac{2}{15}\sqrt{\frac{\pi}{7}}D^X_3\,.
\end{align}
$a^X_{10}$, $a^X_{20}$ and $a^X_{30}$ respectively are the so-called dipole, quadrupole and octupole. For the reader's convenience we give two values of $\lambda_s$: $\lambda_s(z=2)\simeq 5300\,{\rm Mpc}$ and $\lambda_s(z=1)\simeq 3400\,{\rm Mpc}$.

\subsection{Effect on the CMB temperature fluctuations}

Now we review the effects on CMB temperature fluctuations, which in our notation reads \cite{Durrer:2020fza}
\begin{align}\label{eq:CMBDT}
\Delta^{\rm CMB}_{T}(z_{\rm dec},\hat n)=&\frac{1}{4}\delta_{\rm rc}+{\cal H}_{\rm dec}v_{\rm dec}-\delta\omega_{\rm dec}\,.
\end{align}
Note that when computing $\delta\omega_{\rm dec}$ using \eqref{eq:deltaomega} evaluated at decoupling, we have to replace $v_{\rm m}(a_0)$ by $v_o$ as it is our velocity induced by the superhorizon mode. This is not important for adiabatic modes as $v_{\rm r}=v_{\rm m}$ but it is important for isocuvature modes as there is a net relative velocity. Also since the transfer of isocurvature into curvature has not been completed at the time of decoupling, we consider both of their contributions. First, we expand for $\vec{q}\cdot\vec{r}\ll1$ the initial curvature and isocurvature fluctuations as follows
\begin{align}
\Phi_i(\vec{q}\cdot\vec{r})=\sum_{\ell=1}^\infty \frac{1}{\ell!}\Phi_{i,\ell}(qr\cos\theta)^\ell\quad{\rm and}\quad
S_i(\vec{q}\cdot\vec{r})=\sum_{\ell=1}^\infty \frac{1}{\ell!}S_{i,\ell}(qr\cos\theta)^\ell\,.
\end{align}
Note that the wavenumber $q$ of the discrete modes for $\Phi$ and $S$ need not be the same; however, we chose them to be equal for simplicity. Next we take into account the corresponding transfer functions for curvature and isocurvature initial conditions by writing
\begin{align}
\Phi&=\Phi_i(\vec{q}\cdot\vec{x}){\cal T}^{\rm ad}_\Phi(a)+S_i(\vec{q}\cdot\vec{x}){\cal T}^{\rm iso}_{\Phi}(a)\,,\\
v_{\rm r}&=\Phi_i(\vec{q}\cdot\vec{x}){\cal T}^{\rm ad}_{v_{\rm r}}(a)+S_i(\vec{q}\cdot\vec{x}){\cal T}^{\rm iso}_{v_{\rm r}}(a)\,,\\
\delta_{\rm rc}&=S_i(\vec{q}\cdot\vec{x}){\cal T}^{\rm iso}_{\delta_{\rm rc}}(a)\,,
\end{align}
where the transfer functions ${\cal T}^{\rm ad/iso}_Y$ with $Y=\{\Phi, v_{\rm r}, \delta_{\rm rc}\}$ are Eqs.~\eqref{eq:Sxi}-\eqref{eq:vrxi} in Appendix~\ref{subsec:perturbations}. Then, we do the same expansion as before but for the CMB temperature fluctuations:
\begin{align}
\Delta^{\rm CMB}_{T}=\sum_{\ell=1}^\infty \frac{1}{\ell!}D^{\rm CMB}_\ell(\cos\theta)^\ell=\sum_{\ell=1}^\infty a^{\rm CMB}_{\ell0}Y_{\ell0}(\theta,\varphi)\,.
\end{align}
Using the expression for $\delta\omega$ \eqref{eq:deltaomega}, and evaluating at the time of decoupling, we find
\begin{align}\label{eq:dipolequadrupoleoctupole}
D^{\rm CMB}_1&\simeq-0.18(qr_{\rm dec})S_{i,1}\,,\\
D^{\rm CMB}_2&\simeq-(qr_{\rm dec})^2\left(0.32\Phi_{i,2}+0.24S_{i,2}\right)\,,\label{eq:dipolequadrupoleoctupole2}\\
D^{\rm CMB}_3&\simeq-(qr_{\rm dec})^3\left(0.33\Phi_{i,3}+0.24S_{i,3}\right)\,.
\label{eq:dipolequadrupoleoctupole3}
\end{align}
We see that only the initial isocurvature may affect the CMB dipole. The integrals were done analytically. The coefficients of the adiabatic components in Eqs.~\eqref{eq:dipolequadrupoleoctupole2} and \eqref{eq:dipolequadrupoleoctupole3} are in agreement with Ref.~\cite{Erickcek:2008jp}.  However, the coefficients of the isocurvature components are about half of those calculated in Ref.~\cite{Erickcek:2009at}, the likely reason for the discrepancy being that these authors assume the MD era to begin only at CMB decoupling. Hence the curvature perturbation has only reached $\Phi\sim 0.13 S_i$, compared to $\Phi=S_i/5$ for the usual longer period of matter-domination. We believe that this suppression because of an incomplete transfer from isocurvature to curvature accounts for the missing factor of $\sim2$. For the reader's convenience we give here the value for the comoving distance to decoupling:  $r_{\rm dec}=\lambda_{\rm dec}=\eta_0-\eta_{\rm dec}\simeq 14100\,{\rm Mpc}$.

\section{Superhorizon modes and the dipole tension\label{sec:dipoletension}}

Now we can discuss the viability of superhorizon perturbations to alleviate the dipole tension. (The effects of slightly subhorizon modes are discussed in \S~\ref{sec:subhorizonmodes}.) Let us first review measurements of the CMB and quasar number count dipoles. The former has been measured by Planck \cite{Planck:2018nkj} to be
\begin{align}
\label{eq:CMBdipoledata}
D^{\rm CMB}_{1}=(1.23357 \pm 0.00036) \times 10^{-3}\,. 
\end{align}
If it is entirely  kinematic in origin, then we have a precise measurement of the velocity of the Solar system barycentre with respect to the CRF:
\begin{align}
\label{eq:dipole369}
n_iv^i_{o}=369.82\pm0.11~{\rm km/s}
\end{align}
Planck  has also looked for the aberration and modulation effects due to the Observer's velocity \cite{Planck:2013kqc}, obtaining
\begin{align}
\label{eq:aberrationerror}
n_iv^i_{o}=384\pm78\,{\rm (stat)}\pm 115\,{\rm (syst)}\,{\rm km/s},
\end{align}
which is consistent with the kinematic interpretation \eqref{eq:dipole369} but has a large uncertainty. Hence the relative velocity with the CRF can in principle be larger if there is also a significant intrinsic contribution to the CMB dipole. However even when such an intrinsic dipole is allowed to be non-zero, the peculiar velocity of the Solar system inferred from Planck data is $
n_iv^i_{o}=300^{+111}_{-93}\,{\rm km/s}$   \cite{Ferreira:2020aqa,Ferreira:2021omv}. Analysis of the dipolar spectra of the bipolar spherical harmonics of Planck data also finds consistency with the purely kinematic interpretation \eqref{eq:dipole369} at high significance \cite{Saha:2021bay}.

On the other hand the quasar number count dipole is \cite{Secrest:2020has}
\begin{align}
\label{eq:dipolenumberQSOs}
d_{\cal N}=(15.54\pm 1.7)\times 10^{-3}\,,
\end{align}
where the error estimate is from Ref.~\cite{Dalang:2021ruy}. Since we have shown (\S~\ref{sec:superhorizon}) that superhorizon modes do not induce a significant dipole in the number count, let us consider the possibility that Eq.~\eqref{eq:dipolenumberQSOs} is the real kinematic dipole, due to our motion with respect to the CRF of 
\begin{align}
\label{eq:velocityQSOs}
n_iv^i_{o}=(2.66\pm 0.29)\times 10^{-3} \Rightarrow 797\pm87\,{\rm km/s}\,,
\end{align}
This is discrepant at $4.9\sigma$ with the velocity extracted from the CMB if its dipole is purely kinematic in origin. Note however that the velocity \eqref{eq:velocityQSOs} is discrepant  with the value \eqref{eq:aberrationerror} obtained by Planck observations of aberration and modulation at only $3\sigma$.

\begin{figure}
\includegraphics[width=0.7\columnwidth]{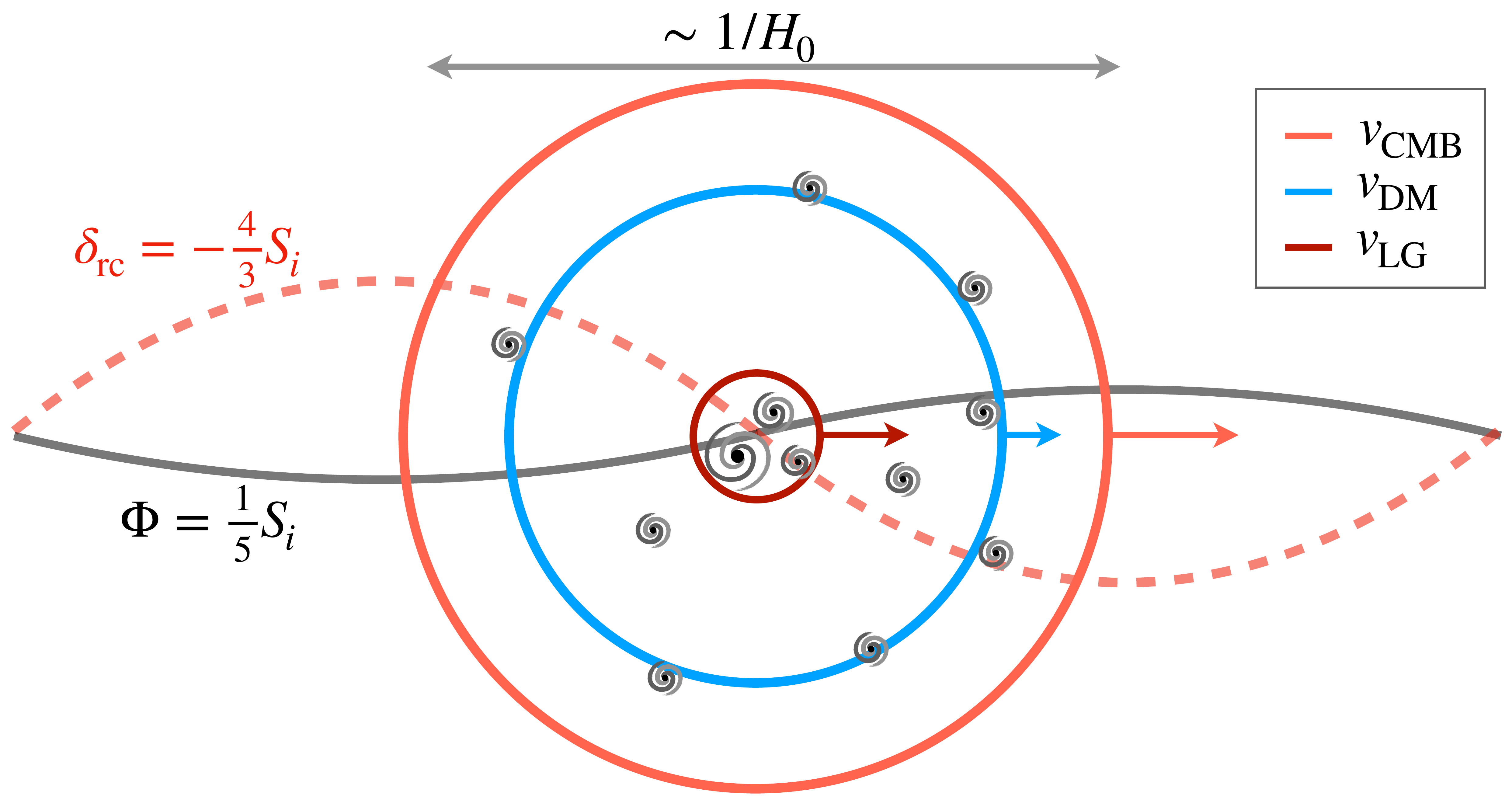}
\caption{Illustration of the effects of an early CDM isocurvature mode today, in the shear-free gauge for easier comparison with the calculations in \S~\ref{sec:superhorizon}. We denote by $v^i_{\rm LG}$, $v^i_{\rm DM}$ and $v^i_{\rm CMB}$, respectively, the velocities of the Local Group, the dark matter rest frame and the CMB rest frame. From the galaxy number count dipole our velocity $v^i_\odot$ with respect to the dark matter rest frame is $n_iv^i_\odot\simeq800\,{\rm km/s}$. However, the CMB rest frame moves away from the matter rest frame with $n_iv^i_{\rm rel}\simeq 800\,{\rm km/s}$, for the parameters in Eq.~\eqref{eq:Siintrinsic}. This, together with the large density fluctuations in the photon field, leads to a CMB dipole consistent with observations. Note that the isocurvature mode needs to be aligned with the CMB dipole. 
\label{fig:illustration}}
\end{figure}

However, if Observer's velocity is as large as Eq.~\eqref{eq:velocityQSOs}, the CMB dipole must have a large (negative) intrinsic contribution in order to be consistent with Eq.~\eqref{eq:CMBdipoledata}. As discussed earlier in \S~\ref{sec:superhorizon}, a superhorizon isocurvature mode can contribute to the dipole \cite{Langlois:1995ca,Langlois:1996ms,Erickcek:2008jp}. Including the effects of such a  mode \eqref{eq:dipolequadrupoleoctupole} we have 
\begin{align}
\label{eq:dipoleCMBtotal}
d^{\rm CMB}=d^{\rm CMB}_{\rm kin}+D^{\rm CMB}_1=n_iv^i_{o}-0.18(qr_{\rm dec})S_{i,1}=1.23357\times10^{-3}\,,
\end{align}
if we assume that the superhorizon mode is aligned with the CMB kinematic dipole, for which there is no obvious rationale. Note also that we consider a single isocurvature mode, which may be thought of as a sharp peak in the power spectrum of isocurvature fluctuations; if the spectrum were flat, we would be constrained by the effect of isocurvature on other CMB multipoles \cite{Planck:2018jri}.\footnote{{In reality, any dynamics that generates such a large superhorizon isocurvature mode would do so over a range of comoving scales with a width that reflects the underlying physical mechanism that generated it. We will return to this issue in a future study where we consider relevant inflationary model constructions.}}
With these important assumptions, we use the mean $v_o$ from Eq.~\eqref{eq:velocityQSOs} to conclude that a superhorizon isocurvature mode with amplitude 
\begin{align}
\label{eq:Siintrinsic}
(qr_{\rm dec})S_{i,1}\simeq 7.9\times 10^{-3}-0.015\left(1-\frac{n_iv^i_{o}}{797~{\rm km/s}}\right)\,,
\end{align}
would yield a CMB dipole compatible with observations. The values above  are possible if, \textit{e.g.}, $q\sim 0.1/r_{\rm dec}\simeq 7\times 10^{-6}\,{\rm Mpc}^{-1}$ and $S_{i,1}\sim 0.1$. Note that while such an early isocurvature superhorizon mode may allow for a large peculiar velocity to be compatible with the CMB dipole, it does not explain why our Solar system moves at $797\,{\rm km/s}$ with respect to the CRF, nor why the isocurvature  mode is aligned with the kinematic dipole. The necessary alignment would have to be accidental, however such a large local velocity with respect to the CRF may be realisable in models where scale-dependent non-Gaussianity is sufficiently enhanced at small comoving scales, but suppressed at larger scales so as to be unconstrained by CMB observations.

We now proceed to investigate the compatibility of this scenario with the CMB quadrupole and study possible connections to future observations. 

\subsection{CMB quadrupole}

Superhorizon modes can be constrained by measuring the CMB quadrupole \cite{Erickcek:2008jp}. Let us consider for simplicity only an early isocurvature mode, so that $\Phi_i=0$. Then we have from the intrinsic dipole \eqref{eq:Siintrinsic} that the induced CMB quadrupole \eqref{eq:dipolequadrupoleoctupole2} is 
\begin{align}
\label{eq:quadrupolecmb}
D^{\rm CMB}_2&\simeq-0.24(qr_{\rm dec})^2S_{i,2}\simeq -1.5\times 10^{-5}\frac{S_{i,2}}{S_{i,1}^2}\,.
\end{align}
Thus, assuming that $S_{i,2}\sim S_{i,1}^2$, as occurs for instance in curvaton models \cite{Erickcek:2009at}, we have
\begin{align}\label{eq:a20}
a_{20}\simeq -8\times 10^{-6}\,.
\end{align}
It is interesting that the Planck measurement  \cite{Planck:2018nkj,Planck:2018vyg} of the quadrupole reads
\begin{align}
{\cal D}_{\ell=2}\simeq 2.3_{-0.9}^{+3.1}\times  10^2\,\mu {\rm K}^2\,,
\end{align}
where we followed their  notation for ${\cal D}_{\ell}$ 
\cite{Schwarz:2015cma,Planck:2018nkj}.
This means that $a_{20}<1.3\times10^{-5}$at $3\sigma$. For comparison the power-law $\Lambda$CDM model best fit \cite{Planck:2018vyg} gives ${\cal D}_{\ell=2}=1016~\mu {\rm K}^2$, \textit{i.e.} the observed quadrupole is  much smaller than the standard model expectation. Hence not only does Eq.~\eqref{eq:a20} provide a contribution  within $3\sigma$ of the observed value, due to the minus sign it can potentially reconcile the larger best-fit expectation with observations. However to make a quantitative prediction, we need a concrete model for the isocurvature mode which we leave for future work.

\subsection{Anisotropy of the Hubble parameter}

We briefly discuss how the superhorizon mode would also affect the luminosity distance. which in our notation (\S~\ref{sec:review}) is:
\begin{align}
d_L=\hat d_o (1+z)^2\,.
\end{align}
We find that the fluctuations in the luminosity distance read
\begin{align}
\frac{\delta d_L}{d_L}=\Delta_n^x\,,
\end{align}
where $\Delta_n^x$ is given in Eq.~\eqref{eq:deltaNs} (see also Refs.~\cite{Bonvin:2005ps,Bonvin:2006en}). Thus we conclude from Eq.~\eqref{eq:nodipole} that the dipole in the luminosity distance is unaffected by the superhorizon mode at leading order in $q\lambda_s$. However, the next-order effect would be more important for the quadrupole. Eq.~\eqref{eq:quadrupolesincount} shows that for low redshift, that is $\lambda_s{\cal H}_0\ll1)$, the quadrupole is approximately given by
\begin{align}
\label{eq:quadrupolehubble}
D_2^{d_L}\simeq-\frac{4}{15}\left(\frac{q}{{\cal H}_0}\right)^2S_{i,2}\sim -6\times 10^{-6}\frac{S_{i,2}}{S_{i,1}^2}\,.
\end{align}
This constant anisotropy at $z\ll1$ in the Hubble parameter from the quadrupole was first noted in Ref.~\cite{Kasai:1987ap}. Although small, it may be detectable in forthcoming large surveys.

\section{Slightly subhorizon modes and the dipole tension\label{sec:subhorizonmodes}}

In the previous section we have considered the effect of superhorizon modes on the galaxy number count fluctuations and found that, in general, no adiabatic superhorizon mode can source a number count dipole at leading order in $q\lambda$. Now we study the effects of slightly subhorizon modes, by which we mean modes which were superhorizon at the time of CMB decoupling but have wavelength much smaller than the path travelled by the photons from the Source. We thus consider wavenumbers in the range $5\times 10^{-3}\,{\rm Mpc}^{-1}>q>7\times 10^{-5}\,{\rm Mpc}^{-1}$, so that $q\eta_{\rm dec}\ll1$ but $q\eta_0>q\lambda_s\gg1$. Modes with larger wavenumbers, \textit{i.e.} $q>5\times 10^{-3}$, contribute to the so-called clustering dipole, which has been estimated to be of ${\cal O}(10^{-4})$ for the quasar sample considered in Ref.~\cite{Secrest:2020has}.  

Let us first clarify our approach. In contrast to \S~\ref{sec:dipoletension}, we assume as is standard that the CMB dipole is kinematic and take $n_iv^i_\odot\simeq 369\,{\rm km/s}$ \eqref{eq:dipole369}. In the general case, one should also consider the possible contribution from superhorizon modes as in \S~\ref{sec:dipoletension}. Then, to be consistent with the results of Ref.~\cite{Secrest:2020has,Secrest:2022uvx}, the number count dipole must have an additional intrinsic contribution, \textit{viz.}
\begin{align}
\label{eq:ellisformula4}
d_{\cal N}=d^{\rm kin}_{\cal N}+d^{\rm int}_{\cal N}\,.
\end{align}
Using Eqs.~\eqref{eq:ellisformula2}, \eqref{eq:dipole369} and \eqref{eq:dipolenumberQSOs}, we find that the intrinsic contribution should be approximately 
\begin{align}\label{eq:intrinsicdipole}
d^{\rm int}_{\cal N}\simeq 9\times 10^{-3}\,.
\end{align}

For subhorizon modes $qr>1$, so we can expand  directly in terms of spherical harmonics $Y_{l0}$ and spherical Bessel functions $j_l$:
\begin{align}
\label{eq:expansionbessel}
\Phi=\sum_{l=0}^\infty \Phi_l (2l+1)j_l(qr)Y_{l0}(\theta)\,,
\end{align}
and similarly for isocurvature, density and velocity perturbations. Since  we assumed that the subhorizon mode is aligned with the number count dipole, we only have the $Y_{l0}(\theta)$ contributions. The time dependence of $\Phi_l$ is given by the transfer functions of a radiation+matter-dominated universe. Since we are dealing with subhorizon modes, we numerically solve the Einstein equations given in Appendix~\ref{app:perturbations} and set the initial conditions deep inside the radiation-dominated universe. For purely adiabatic modes we can  use the analytical solutions given in Eqs.~\eqref{eq:dustdominatedT} and \eqref{eq:deltamc2}. We do not include the effect of the cosmological constant $\Lambda$ which is only relevant at $z\lesssim0.8$. In general the effect of $\Lambda$ is to suppress structure growth so the estimates below should be understood as a conservative upper bound. Also, we do not consider the effects of baryons so the final results are uncertain to $\sim 10\%$, however this does not change our conclusions. 

\begin{figure}
\includegraphics[width=0.5\columnwidth]{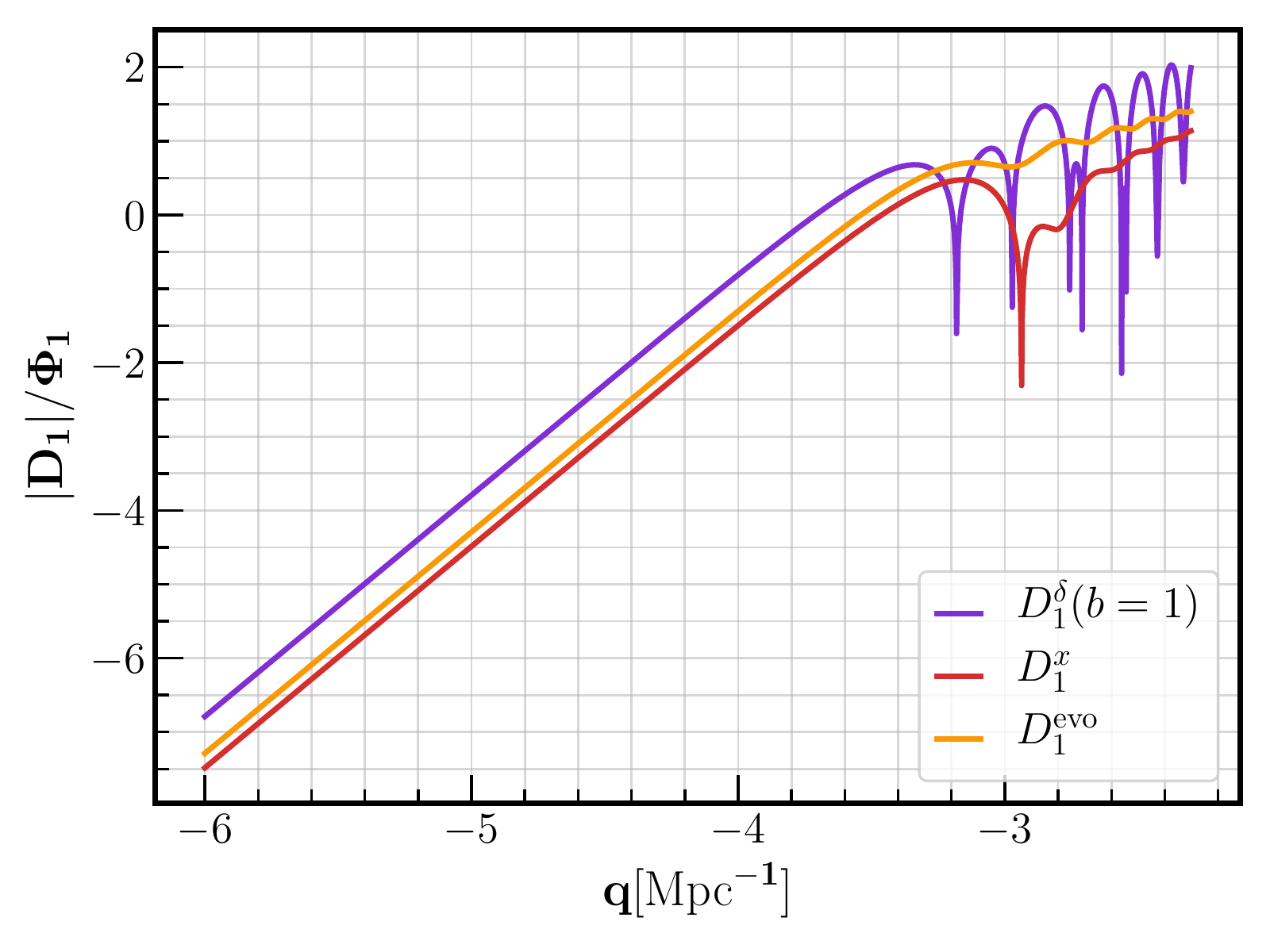}
\caption{Dipole contributions to the galaxy number count \eqref{eq:deltaNready} in terms of wavenumber at redshift $z=2$, normalised to the initial amplitude $\Phi_1$. We see that the larger the wavenumber the more subhorizon a mode is, hence the larger the contribution to the dipole,  reaching ${\cal O}(10)$ for $q\sim 10^{-3}\,{\rm Mpc}^{-1}$. The oscillations present for $q\gtrsim 10^{-3}\,{\rm Mpc}^{-1}$ are due to the Taylor expansion \eqref{eq:expansionbessel} in terms of Bessel functions. \label{fig:subhorizon}}
\end{figure}

\begin{figure}
\includegraphics[width=0.49\columnwidth]{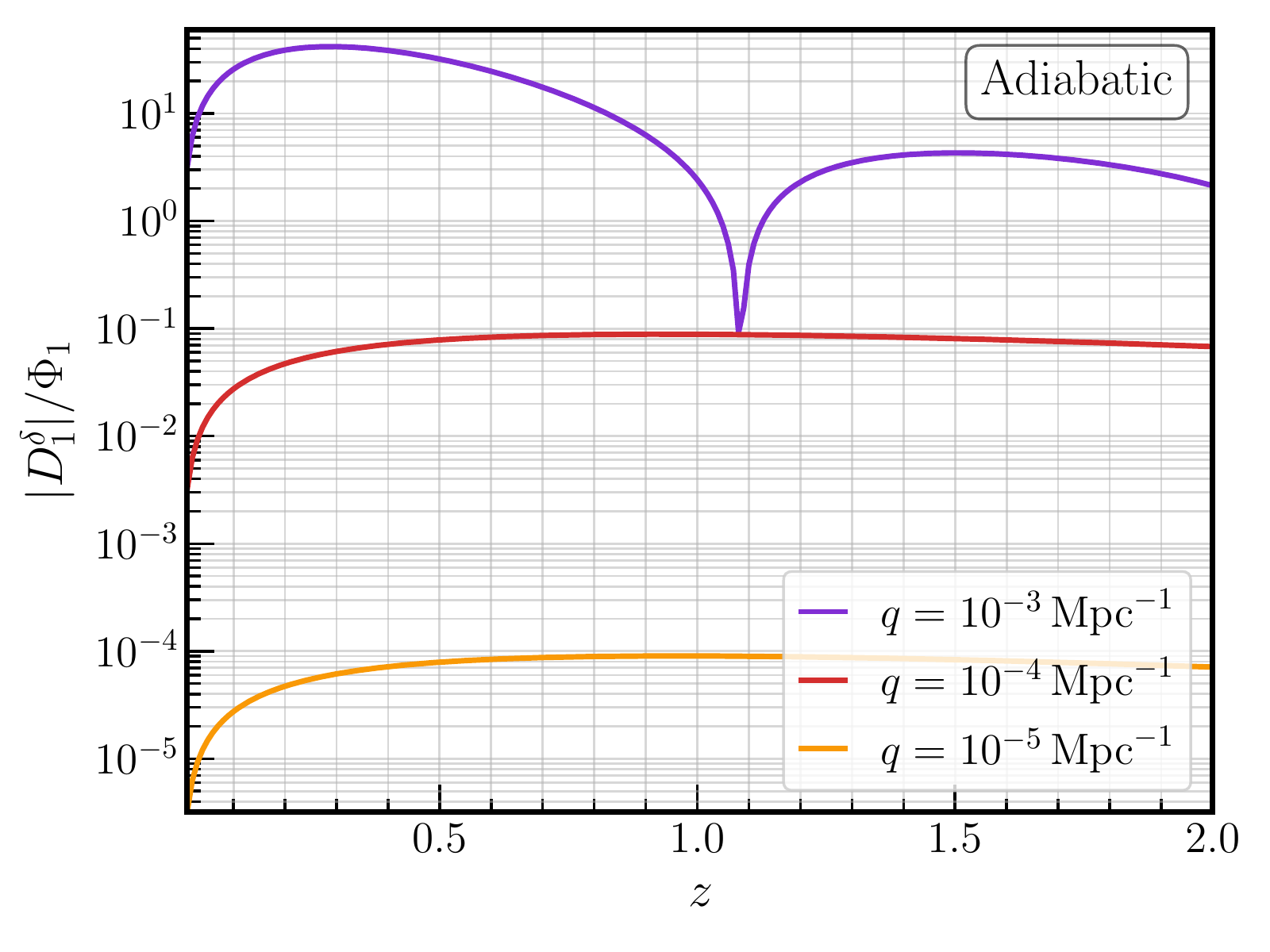}
\includegraphics[width=0.49\columnwidth]{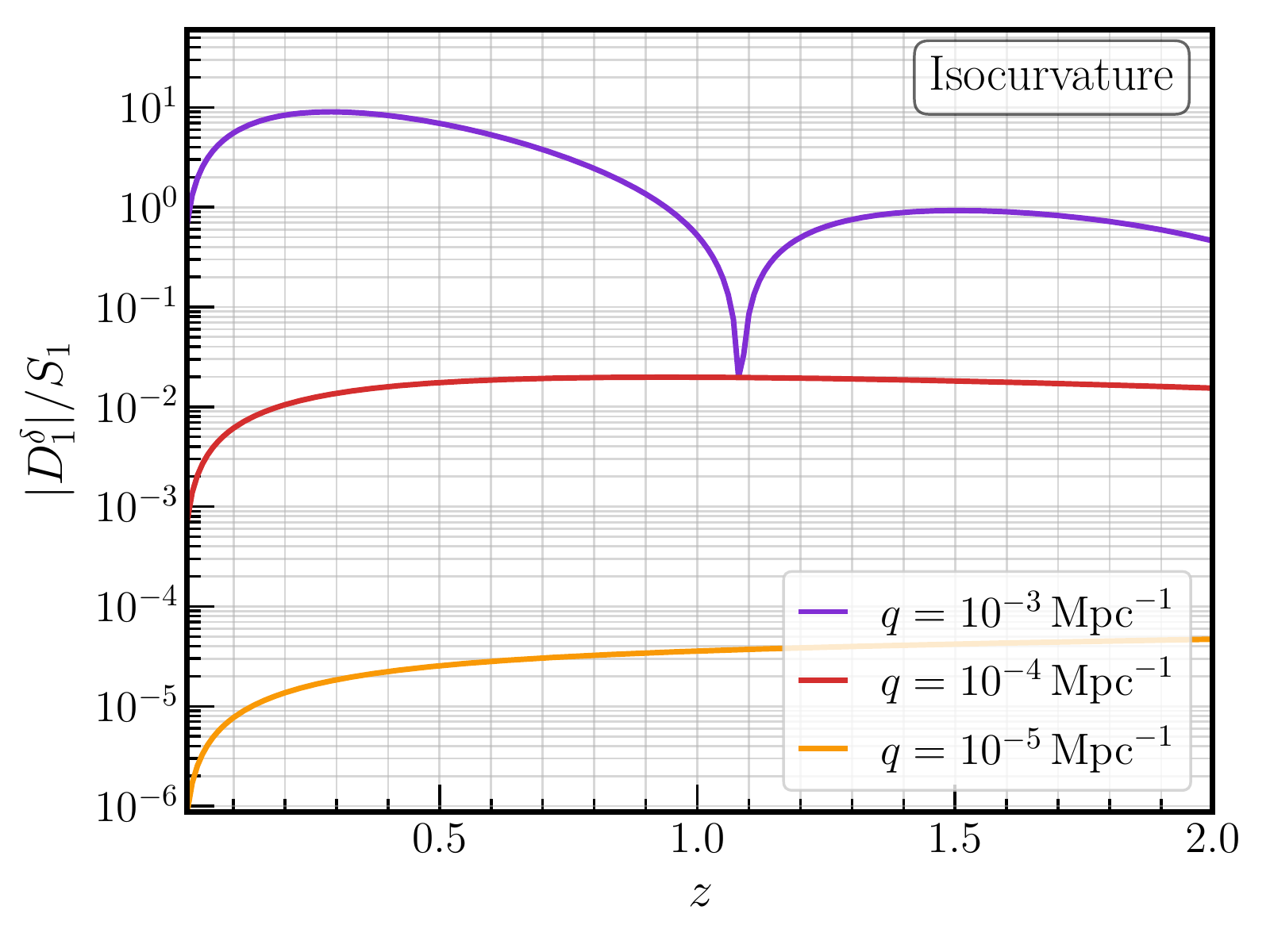}
\caption{Contribution $D_1^\delta$\eqref{eq:deltaN1} to the number count dipole \eqref{eq:deltaNready} versus redshift for modes with $q=\{10^{-3}$ (purple), $10^{-4}$ (red), $10^{-5}$ (orange)$\}~{\rm Mpc}^{-1}$, for adiabatic (left) and isocurvature (right) initial conditions. 
\label{fig:subhorizonnumerics}}
\end{figure}

With the numerical solutions to the radiation+matter-dominated universe and the expansion in terms of Bessel functions \eqref{eq:expansionbessel}, we compute the dipole of the number count using Eq.~\eqref{eq:deltaNready}.  Fig.~\ref{fig:subhorizon} shows the dependence of the different contributions to the dipole in \eqref{eq:deltaNready} with wavenumber $q$ for a pure adiabatic mode. We see that the larger the wavenumber, the more density fluctuations have grown and the larger the dipole. However, the larger the wavenumber, the tighter too are the CMB constraints from Planck \cite{Planck:2018jri}. Roughly we have that $\Phi_i<5.2\times 10^{-5}$ for $q\sim 10^{-3}\,{\rm Mpc}^{-1}$, where $\Phi_i$ is the amplitude of the primordial curvature perturbation. Isocurvature perturbations can have an amplitude of at most $10\%$ of the curvature perturbation, hence $S_i<10^{-5}$ for $q\sim 10^{-3}\,{\rm Mpc}^{-1}$.

\begin{figure}
\includegraphics[width=0.49\columnwidth]{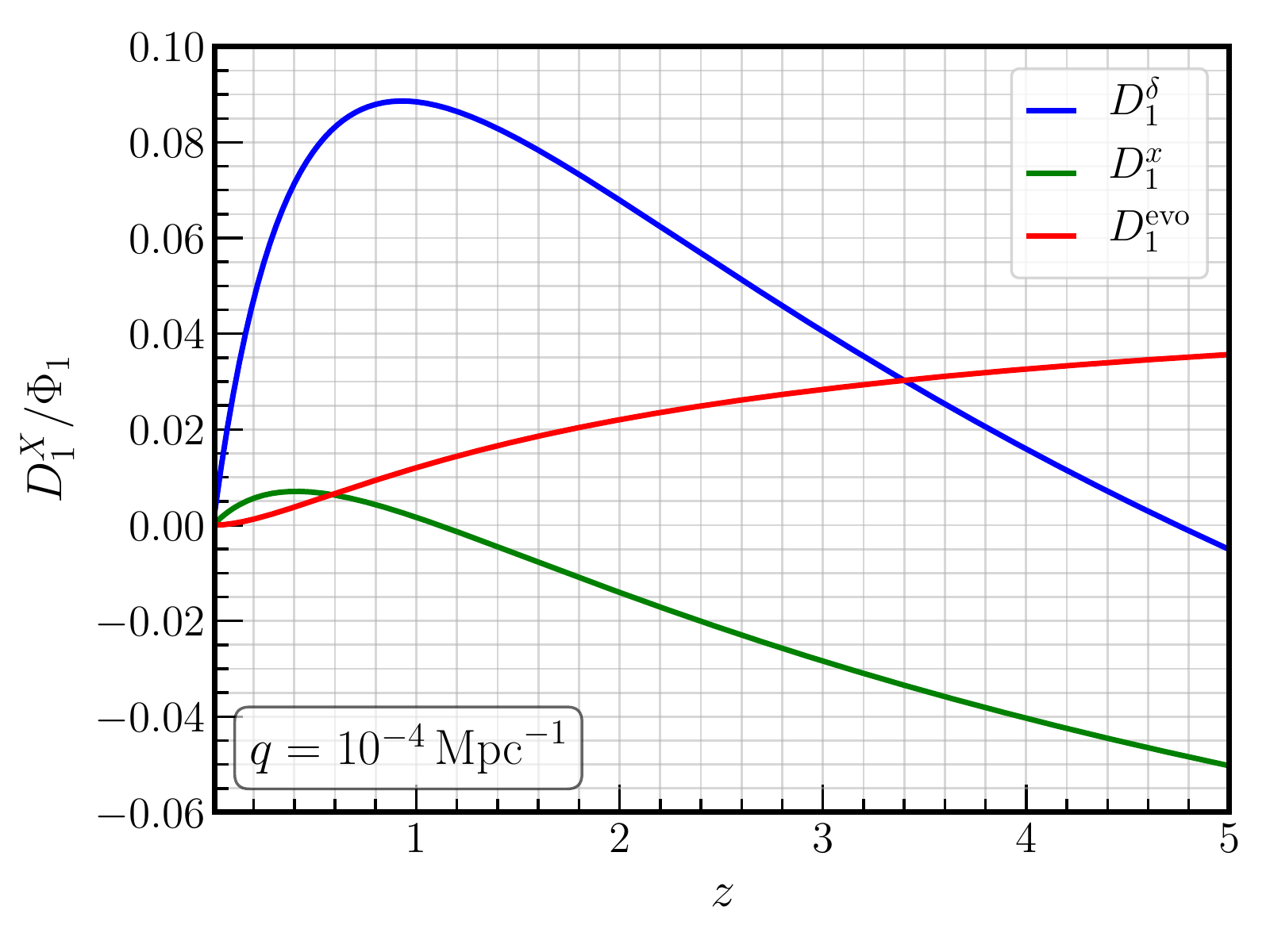}
\includegraphics[width=0.49\columnwidth]{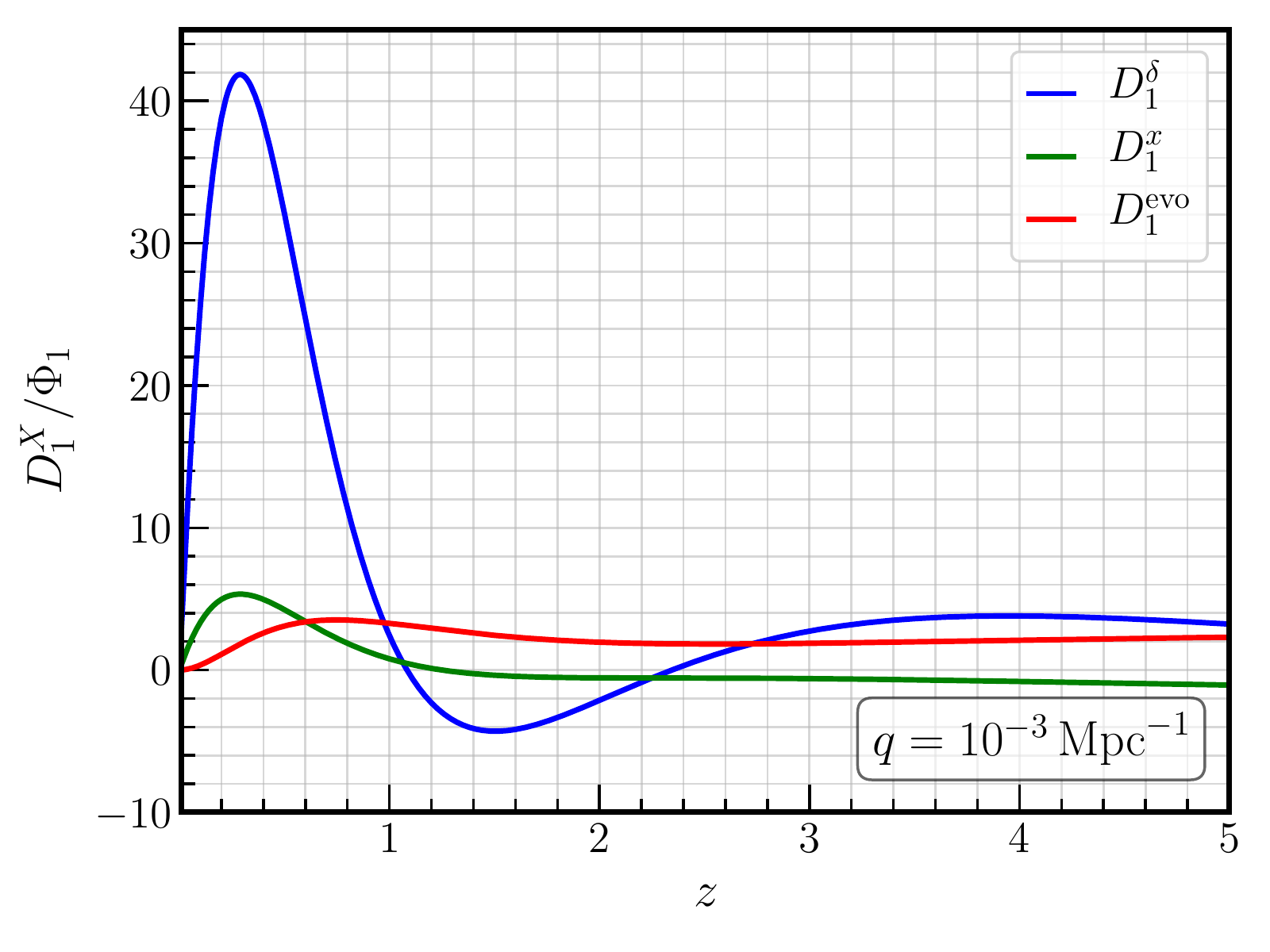}
\caption{Contributions to the number count dipole \eqref{eq:deltaNready} versus redshift for $q=10^{-4}{\rm Mpc}^{-1}$ (left panel) and $q=10^{-3}{\rm Mpc}^{-1}$ (right panel), both for adiabatic initial conditions. For $q=10^{-4}{\rm Mpc}^{-1}$ the largest contribution peaks at $z\sim 1$ with amplitude $D_1^{\delta}\sim0.09\,\Phi_1$. For $q=10^{-3}{\rm Mpc}^{-1}$ the largest contribution peaks at $z\sim 0.5$ with amplitude $D_1^{\delta}\sim42\, \Phi_1$. \label{fig:subhorizonnumerics2}}
\end{figure}

In Fig.~\ref{fig:subhorizonnumerics} we show the redshift dependence of the contribution $D_1^\delta$ \eqref{eq:deltaN1} to the number count dipole \eqref{eq:deltaNready} for wavenumbers $q= \{10^{-3},10^{-4},10^{-5}\}\,{\rm Mpc}^{-1}$ and for both adiabatic and isocurvature initial conditions. We see that the contribution of initial isocurvature to the number count is suppressed by nearly an order of magnitude with respect to the adiabatic case. Hence for simplicity, we proceed with slightly subhorizon adiabatic modes, which are also not as constrained by Planck \cite{Planck:2018jri} as isocurvature fluctuations. The contributions to the number count dipole for $q=10^{-4}{\rm Mpc}^{-1}$ and $q=10^{-3}{\rm Mpc}^{-1}$ for this case are shown in Fig.~\ref{fig:subhorizonnumerics2}. For $q=10^{-4}{\rm Mpc}^{-1}$ we see that at redshifts of $z\sim 1$ the largest contribution comes from $D_1^\delta$ with an amplitude of the order of $0.09\,\Phi_1$. However, taking into account the above-mentioned CMB constraints \cite{Planck:2018jri}, the dipole from $q=10^{-4}{\rm Mpc}^{-1}$ is $D_1^\delta\sim{\cal O}(10^{-5})$. For $q=10^{-3}{\rm Mpc}^{-1}$ we see that $D_1^\delta$ peaks at $z\sim 0.5$ with an amplitude $D_1^\delta\sim 42 \,\Phi_1$. The dipole allowed by CMB constraints is then $D_1^\delta\sim{\cal O}(10^{-3})$. Note, however, that the intrinsic dipole in the total number count is given by
\begin{align}
d^{\rm int}_{\cal N}&=\frac{1}{\int_0^{\lambda_s} {\cal N}_s(\lambda,L) r^2d\lambda}\int_0^{\lambda_s} \left(b\,D_1^\delta-2xD_1^x-f_{\rm evo}D_1^{\rm evo}\right) {\cal N}_s(\lambda,L) r^2d\lambda\,.
\end{align}
Although an accurate calculation requires knowledge of the evolution of the sources in redshift which is uncertain, we can obtain a rough  estimate as follows. First, since for $q\sim10^{-3}{\rm Mpc}^{-1}$, $D_1^\delta$ is by far the dominant contribution at $z<1$ and the quasar numbers also peak at $z\sim 1$, we may evaluate the integrals at this redshift. Then, taking $b=1$, this gives $d^{\rm int}_{\cal N}\sim D_1^\delta$. This implies that a discrete adiabatic mode with $q\sim10^{-3}{\rm Mpc}^{-1}$ can yield at most $d^{\rm int}_{\cal N}\sim {\cal O}(10^{-3})$. Hence this cannot solve the dipole tension but it suggests that a collection of modes with $q\sim10^{-3}{\rm Mpc}^{-1}$, say corresponding to a `bump' in the primordial spectrum (as seen in reconstructions, \textit{e.g.} Figure 2 of Ref.~\cite{Hunt:2015iua}), could add up to a sizeable contribution to the number count dipole. The above order of magnitude estimate illustrates that while slightly subhorizon modes cannot explain the number count dipole,  they do deserve further study which we leave for future work. 

\section{Discussion \& conclusions\label{sec:conclusions}}

\begin{table}
\renewcommand{\arraystretch}{1.5}
\begin{tabularx}{\textwidth}{|X|X|X|}
\toprule[2.0pt]\addlinespace[0mm]
\multicolumn{1}{|c|}{\diagbox{Size of mode q}{Initial conditions}} & \centering {\textbf{Adiabatic discrete mode}}& \multicolumn{1}{|c|}{\textbf{Isocurvature discrete mode}} \\
\toprule\addlinespace[0mm]
\textbf{Superhorizon} \newline ($q<{\cal H}_0$) & No CMB dipole${^*}$ \cite{Erickcek:2008jp}\newline No NC dipole${^*}$ \newline Cannot solve dipole tension
& Intrinsic CMB dipole \cite{Erickcek:2008jp}\newline No NC dipole$^{*}$ \newline Might resolve dipole tension$^{**}$\\
\hline
\textbf{Slightly subhorizon} \newline(${\cal H}_0\lesssim q\lesssim{\cal H}_{\rm dec}$)  & Amplitude $\lesssim8\times10^{-5}$ (CMB \cite{Planck:2018jri})\newline ${\cal O}(10^{-3})$ maximum NC dipole \newline Cannot solve dipole tension & Amplitude $\lesssim 10\%$ of adiabatic \cite{Planck:2018jri} \newline ${\cal O}(10^{-4})$ maximum NC dipole \newline Cannot solve dipole tension\\
\hline
\textbf{Subhorizon}\newline($q\gtrsim {\cal H}_{\rm dec}$) & Amplitude $\sim 5\times10^{-5}$ \cite{Planck:2018jri} \newline Cannot solve dipole tension \cite{Secrest:2020has}& Amplitude $\lesssim 10\%$ of adiabatic \cite{Planck:2018jri} \newline Cannot solve dipole tension \\
\toprule[2.0pt]
\end{tabularx}
\flushleft{$^*$ At leading order in $qr$.\newline
$^{**}$ If our peculiar velocity with respect to the CRF is $\sim 800\,{\rm km/s}$.}
\caption{{\it Summary of the main results of this work:} Adiabatic superhorizon modes do not affect the CMB and number count (NC) dipoles. Early isocurvature superhorizon modes affect only the CMB dipole because after CDM dominates isocurvature modes are converted to adiabatic modes. A sizeable intrinsic CMB dipole from an early isocurvature superhorizon mode can resolve the dipole tension if our peculiar velocity is $\sim800\,{\rm km/s}$. Slightly subhorizon modes are constrained by CMB \cite{Planck:2018jri} but with large errors due to cosmic variance. Slightly subhorizon adiabatic modes with $q\sim 10^{-3}\,{\rm Mpc}^{-1}$ may contribute a NC dipole of ${\cal O}(10^{-3})$, which is not sufficient to explain the dipole tension; however, a more detailed study is needed to rule out this possibility. The current comoving horizon and the comoving horizon at decoupling are given respectively by ${\cal H}_0\simeq 2\times 10^{-4}\,{\rm Mpc}^{-1}$ and ${\cal H}_{\rm dec}\simeq 5\times 10^{-3}\,{\rm Mpc}^{-1}$.\label{tab:table1}}
\end{table}

The recently reported mismatch between the CMB dipole \cite{Planck:2018nkj} and the predicted quasar number count dipole \cite{Secrest:2020has,Secrest:2022uvx} suggests that the rest frames of the CMB and dark matter are different. We have investigated if this tension can be due to the effects of primordial fluctuations with wavelength exceeding or similar to the current Hubble radius. Our  main results are summarised in Table~\ref{tab:table1}.

We find that, as for the CMB dipole \cite{Erickcek:2008jp}, an initial adiabatic superhorizon mode does not affect the number count dipole at leading order in $qr<1$, where $q$ is the wavenumber of the superhorizon mode and $r$ the comoving distance travelled by the photons. Next, we considered initial CDM  isocurvature perturbations, which may originate in axion or curvaton models. However, since isocurvature perturbations are converted into adiabatic ones in the matter-dominated era \cite{Kodama:1986ud}, we find that there is practically no induced dipole in the number count of sources which are mainly at redshifts $z\lesssim 2$. Therefore initial CDM superhorizon isocurvature perturbations do not affect the number count dipole either. However, these can induce an intrinsic CMB dipole \cite{Turner:1991dn,Langlois:1995ca,Erickcek:2009at}.

We thus argued that an isocurvature superhorizon mode with an amplitude of $\sim0.1$ and a wavelength of $\sim10$ times the current horizon can reconcile the quasar number count \cite{Secrest:2020has,Secrest:2022uvx} and the CMB dipoles \cite{Planck:2018nkj} by allowing for a large peculiar velocity of the Solar system of about 800~km/s with respect to the CRF. The isocurvature superhorizon mode provides a \textit{negative} intrinsic contribution to the CMB dipole, cancelling about half of the kinematic dipole due to our large peculiar velocity. The required amplitude is compatible with measurements of the CMB quadrupole. 

While an early isocurvature superhorizon mode  allows for a large peculiar velocity to be compatible with the CMB dipole, it does not explain why the intrinsic dipole should be aligned with the CMB dipole, nor why our Solar system should move at $\sim 800$~km/s with respect to the CRF.\footnote{Our local bulk flow has been reported to extend to $\sim200h^{-1}$\,Mpc and beyond with velocity of $\sim200-250$\,km/s \cite{Colin:2010ds,Ma:2010ps,Dai:2011xm,Feindt:2013pma,Appleby:2014kea,Stahl:2021mat,Abdalla:2022yfr}. The Gpc-sized fluctuations that we are considering in this paper are not relevant for such small-scale bulk flows unless these extend out to cosmological scales as was claimed in  Ref.\cite{Kashlinsky:2008ut}.} Nearby superclusters can account for up to $\sim300\,{\rm km/s}$ of that peculiar velocity (see {\it e.g.} \citep{Colin:2010ds} and references therein). The remaining $\sim500\,{\rm km/s}$ might be due to even bigger structures beyond our Local Group or due to {large non-Gaussianities on small scales}. We note, however, that the presence of such large inhomogeneities beyond $200/h\,{\rm Mpc}^{-1}$ is not expected in $\Lambda$CDM and may be in conflict with CMB lensing observations. Such potential ways to falsify the superhorizon scenario and the investigation of the impact of large non-Gaussianities on small scales on the local velocity dispersion are beyond the scope of this paper. But we plan to explore them in subsequent works. 


As an alternative, we explored in \S~\ref{sec:subhorizonmodes} the effect of slightly subhorizon modes, \textit{i.e.,} subhorizon today but superhorizon at decoupling, now assuming that the CMB dipole is kinematic but that there may be an intrinsic contribution to the number count dipole. First, we find that the effect of initial adiabatic fluctuations on the dipole is larger than the early isocurvature case. Second, we find that since both CDM density fluctuations and velocities of subhorizon modes grow with time, the dipole due to such fluctuations is larger at larger wavenumbers and smaller redshifts. For instance, the dipole for $q\sim 10^{-3}\,{\rm Mpc}^{-1}$ peaks at $z\sim 0.5$ with an amplitude of $\sim 40$ times the amplitude of primordial fluctuations. However, CMB measurements constrain the amplitude of the primordial fluctuations to be $\Phi_i<5\times 10^{-5}$. Thus, discrete slightly subhorizon adiabatic modes with $q\sim 10^{-3}\,{\rm Mpc}^{-1}$ can at most generate a dipole in the number count of ${\cal O}(10^{-3})$, which is insufficient to explain the large number count dipole \cite{Secrest:2020has,Secrest:2022uvx}. Nevertheless, there is the possibility that a collection of modes with $q\sim 10^{-3}\,{\rm Mpc}^{-1}$ may generate a sizeable dipole. A characteristic signal of such slightly subhorizon modes is a redshift dependent dipole with a peak at $z\sim 0.5$. We leave a detailed study of this possibility for future works.

We have neglected the effects of baryons and neutrinos, however this does not significantly affect our results. Furthermore, in the late universe baryons fall into the dark matter gravitational potential wells and therefore follow the CDM  distribution. Therefore we do not expect that baryons or initial baryon isocurvature would change our conclusions concerning the galaxy number count. An interesting possibility worth studying is the effect of initial velocity isocurvature perturbations (or any process that can induce relative velocities), although it is not clear how such perturbations can be generated in the early universe. Moreover for simplicity we focused on a single discrete CDM isocurvature mode. Hence our analysis should apply to isocurvature perturbations with a sharp peak in their primordial spectrum, but further work is required for a broad featureless spectrum. Also we have not presented any  concrete model that can generate such isocurvature modes. The simplest assumption is that these are remnants of the pre-inflationary universe \cite{Langlois:1996ms,Langlois:1996rx,Garcia-Bellido:1997tjw,Linde:1998iw,Linde:1999wv}. It would be interesting also to consider the above scenario in axion isocurvature models. 

The superhorizon isocurvature scenario proposed in this work to resolve the dipole tension will be tested by future experiments. First, proposed future CMB experiments \cite{CORE:2017krr} and new techniques \cite{Meerburg:2017xga,Yasini:2019ajn} will test the kinematic hypothesis more precisely. For instance, a (non)-detection of a CMB intrinsic component of ${\cal O}(10^{-3})$ could (rule out) hint towards an isocurvature superhorizon mode; for recent work in this direction see \cite{Ferreira:2020aqa,Ferreira:2021omv}. Also, an intrinsic CMB dipole component implies a frequency dependence of the CMB quadrupole \cite{Kamionkowski:2002nd}. Second, information on the redshift dependence of the number count dipole as well as possibly the number count quadrupole will come from the Legacy Survey of Space \& Time (LSST) by the Rubin Observatory \cite{LSSTScience:2009jmu}, as well the Square Kilometre Array (SKA)  \cite{Bengaly:2018ykb}. The isocurvature superhorizon mode scenario implies a constant (purely kinematic) dipole in the galaxy number count, and a quadrupole which is redshift dependent but approaches a constant today. The quadrupole in the total number count due to the superhorizon mode is predicted to be of ${\cal O}(10^{-6})$. Such a quadrupole \eqref{eq:quadrupolehubble} would also appear in the Hubble parameter \cite{Kasai:1987ap} although this would be hard to detect.  

The nature of the dipole may also be probed by other observations, such as the cosmic infrared background \cite{Fixsen:2011qk,Kashlinsky:2022tit}, CMB spectral distortions \cite{Chluba:2016bvg}, the stochastic gravitational wave background \cite{Kashyap:2022ibx,Galloni:2022rgg,DallArmi:2022wnq} and the 21-cm line \cite{Slosar:2016utd}. It would also be interesting to perform a general joint analysis of CMB and the number counts dipole, without making any kinematic assumption.

We conclude by noting that the inference of a Cosmological Constant $\Lambda$ from late-universe observations assumes that galaxies are isotropically distributed in the reference frame in which the CMB is isotropic. Since this is apparently not the case with $>5\sigma$ significance \cite{Secrest:2020has,Secrest:2022uvx}, even if
the dipole tension is resolved in a FLRW framework with superhorizon fluctuations as we have discussed, it poses a serious challenge to the standard $\Lambda$CDM cosmology. 


\section*{Acknowledgments} 
G.D. would like to thank M.~Sasaki for useful discussions and D.~Bertacca, M.~Kamionkowski, S.~Matarrese, A.D.~Rojas and D.~Rojas for helpful comments. R.M. \& S.S. thank J.~Peebles for interesting remarks and encouragement. G.D. was supported by the European Union's Horizon 2020 research and innovation programme as a Fellini fellow under the Marie Sk{\l}odowska-Curie grant agreement No 754496.

\appendix

\section{Perturbative expansion in the number count \label{app:perturbativenumber}}

In this Appendix we provide the details necessary to follow the calculations in \S~\ref{sec:review}.

\subsection{Photon's geodesics}
After expansion of Eq.~\eqref{eq:nullgeodesics}, the perturbed photon's geodesic is given by
\begin{align}
\frac{d \delta  k^\mu}{d\bar \lambda}=-\delta\Gamma_{\alpha\beta}^\mu  k^\alpha  k^\beta\,,
\end{align}
where
\begin{align}
\delta\Gamma_{\mu\nu}^\alpha=2\delta^\alpha_0\delta^0_{(\mu}\partial_{\nu)}\Psi+2\delta^i_{(\mu}\partial_{\nu)}\Phi\delta^\alpha_{i}+\eta^{\alpha\lambda}\partial_\lambda \Psi\delta^0_{(\mu}\delta^0_{\nu)}-\eta^{\alpha\lambda}\partial_\lambda\Phi\delta^i_{(\mu}\delta^j_{\nu)}\delta_{ij}\,.
\end{align}
Contracting with the null vectors, we arrive at
\begin{align}\label{eq:deltak0}
\frac{d  \delta  k^0}{d\lambda}=2\frac{d\Psi}{d\lambda}+\frac{\partial}{\partial \eta}(\Psi-\Phi)\,,
\end{align}
and
\begin{align}
\frac{d  \delta  k^i}{d\lambda}=-2n^i\frac{d\Phi}{d\lambda}-\partial^i(\Psi-\Phi)\,.
\end{align}

\subsection{Observer's area distance}
Here we solve the equations for the Observer's area distance $\tilde d_0$ using \eqref{eq:areachange} and \eqref{eq:geodesicdeviation}. First, we have that in the background
\begin{align}
\theta=\partial_i n^i=\frac{2}{r(\lambda)}\quad{\rm and}\quad r(\lambda)=\lambda.
\end{align}
since $n^i$ is a vector in the radial direction $r$ from us. The latter equality holds in a flat FLRW model \cite{Kasai:1987ap}. Then, at first-order we consider the variables $\tilde d_o=r(1+\delta_r)$ and $\tilde\theta=2/r(1+\delta_\theta)$, with which Eqs.~\eqref{eq:areachange} and \eqref{eq:geodesicdeviation} become
\begin{align}\label{eq:dr}
\frac{d}{d\lambda}\delta_r&=\frac{1}{r(\lambda)}\delta_\theta\,,\\
\frac{d}{d\lambda}\left(r(\lambda)\delta_\theta\right)&=-\frac{r^2(\lambda)}{2} \tilde R_{\mu\nu} k^\mu  k^\nu\,.\label{eq:dtheta}
\end{align}
We have also used that since $\tilde\sigma_{\mu\nu}\tilde\sigma^{\mu\nu}=\tilde\nabla_\mu \tilde k^\nu\tilde \nabla_\nu \tilde k^\mu-\tilde\theta^2/2$, at the background level we have $\sigma_{\mu\nu}\sigma^{\mu\nu}=0$. This implies that $\sigma_{\mu\nu}$ starts at first-order in perturbation theory and that $\sigma_{\mu\nu}\sigma^{\mu\nu}$ is a second-order quantity. After integration of Eqs.~\eqref{eq:dr} and \eqref{eq:dtheta} we find
\begin{align}
\delta_r(\lambda_s)=\delta_r(0)-\frac{1}{2}\int_0^{\lambda_s}\frac{d\lambda_1}{r^2(\lambda_1)}\int_0^{\lambda_1}d\lambda_2 \,r^2(\lambda_2) \tilde R_{\mu\nu} k^\mu  k^\nu\,.
\end{align}

Expanding the Ricci tensor at linear order in perturbation theory, we obtain
\begin{align}\label{eq:rmunukmuknu}
\tilde R_{\mu\nu} k^\mu k^\nu=\partial_\alpha\partial^\alpha\left(\Psi-\Phi\right)-2\frac{d}{d\lambda}\frac{\partial\Psi}{\partial\eta}+2\frac{d}{d\lambda}n^i\partial_i\Phi-\frac{d^2}{d\lambda^2}\left(\Psi+3\Phi\right)\,,
\end{align}
where we used that at leading order $dn^i/d\lambda=0$. 
In the above equation, we used that in cartesian coordinates $\tilde\Gamma_{\mu\nu}^\alpha$ is a first-order quantity and therefore at leading order
\begin{align}
\tilde R_{\mu\nu}=\partial_\alpha\tilde\Gamma_{\mu\nu}^\alpha-\partial_\mu\partial_\nu\ln\sqrt{-\tilde g}
\end{align}
Then we also used that
\begin{align}
\partial_\mu\partial_\nu\ln\sqrt{-\tilde g}=\tilde\nabla_\mu\tilde\nabla_\nu\ln\sqrt{-\tilde g}+\tilde\Gamma_{\mu\nu}^\alpha\tilde\nabla_\alpha\ln\sqrt{-\tilde g}
\end{align}
so that at leading order
\begin{align}
\tilde k^\mu \tilde k^\nu\tilde\nabla_\mu\tilde\nabla_\nu\ln\sqrt{-\tilde g}=\tilde k^\mu \tilde\nabla_\mu(\tilde k^\nu\tilde\nabla_\nu\ln\sqrt{-\tilde g})-\tilde k^\mu \tilde\nabla_\mu\tilde k^\nu\tilde\nabla_\nu\ln\sqrt{-\tilde g}=\frac{d^2}{d\lambda^2}\ln\sqrt{-\tilde g}\,,
\end{align}
where we used the geodesic equation $\tilde k^\mu\tilde\nabla_\mu \tilde k^\nu=0$. After some simplification, we can write Eq.~\eqref{eq:rmunukmuknu} as
\begin{align}
\tilde R_{\mu\nu} k^\mu k^\nu=-\frac{1}{r^2}\Delta_\Omega(\Psi-\Phi)+\frac{d^2}{d\lambda_2^2}\Phi-\frac{2}{r}\partial_r(\Psi-\Phi)\,.
\end{align}

\subsection{Redshift}

Here we provide details on the calculation of $\delta\lambda_s$ to derive the $N-z$ relation. We start with the energy of the photon at a given $\bar\lambda$ and $\hat n$, which is given by
\begin{align}
\hat\omega=(- \hat k_\mu \hat u^\mu)=a^{-1}(\tilde\eta)(\tilde k_\mu \tilde u^\mu)=a^{-1}(\eta)(1+\Psi+n_iv^i-\delta k^0-{\cal H}\delta\eta)\,.
\end{align}
However, we observe at a fixed redshift, not at a fixed affine parameter. So let us consider $\lambda=\lambda_s+\delta\lambda_s$ where $\lambda_s=\eta_o-\eta_s$. Then
\begin{align}
1+\tilde z(\lambda_s+\delta\lambda_s)=1+z(\lambda_s)=\frac{a(\eta(0))}{a(\eta(\lambda_s))}\,.
\end{align}
Expanding, we find that
\begin{align}
1+\tilde z(\lambda_s)=1+z(\lambda_s-\delta\lambda_s)=1+z(\lambda_s)-\frac{\partial z}{\partial \lambda_s}\delta\lambda_s\,.
\end{align}
Solving for $\delta\lambda_s$, we obtain
\begin{align}
\delta\lambda_s=\frac{1}{{\cal H}(\lambda_s)}\big[{\cal H}\delta\eta-\delta\omega\big]_{0}^{\lambda_s}\,,
\end{align}
where we integrated Eq.~\eqref{eq:deltak0} once and then twice, which yields
\begin{align}
\delta  k^0(\lambda_s)= \delta k^0(0)+2\left(\Psi(\lambda_s)-\Psi(0)\right)+\int_0^{\lambda_s}d\lambda\frac{\partial}{\partial \eta}(\Psi-\Phi)\,.
\end{align}
and
\begin{align}
\delta\eta=\int_{0}^{\lambda_s}d\lambda\,\delta k^0(\lambda,\hat n)\,.
\end{align}

We set the conditions at the Observer position by requiring that $\delta\omega(\lambda\to0)\to 0$ and $\delta_r(\lambda\to 0)$, i.e. we do not see any change if we are on top of the Source. By doing so, we have:
\begin{align}
\delta k^0(0)=\Psi(0)+n_iv^i_o \quad {\rm and}\quad
\delta_r(0)=\Phi(0)\,.
\end{align}
With these two last equalities, our calculation provides the galaxy number count including dipole and monopole contributions at the Observer.

\section{Comparison with other works \label{app:comparison}}

In this Appendix we show that our formula \eqref{eq:deltaNready} is equivalent to that of Refs.~\cite{Bonvin:2011bg,Challinor:2011bk}. First, we rewrite Eqs.~\eqref{eq:deltaeta} and \eqref{eq:deltar} after doing integration by parts as
\begin{align}
\label{eq:deltaetaapp}\delta\eta&=\lambda n_iv^i_o+\int_{0}^{\lambda}d\lambda_1\left(\Psi+\Phi-\Psi_o\right)+\int_{0}^{\lambda}d\lambda_1(r(\lambda)-r(\lambda_1))n^i\partial_i(\Psi-\Phi)\,,
\end{align}
and
\begin{align}\label{eq:deltarapp}
\delta_r(\lambda)=\Phi_o+\Phi&-\frac{2}{r(\lambda)}\int_0^\lambda d\lambda_1\Phi-\frac{1}{2r(\lambda)}\int_0^{\lambda}d\lambda_1\frac{r(\lambda)-r(\lambda_1)}{r(\lambda_1)}\Delta_\Omega(\Psi-\Phi)\nonumber\\&-\frac{1}{r(\lambda)}\int_0^{\lambda}d\lambda_1(r(\lambda)-r(\lambda_1))n^i\partial_i(\Psi-\Phi)\,,
\end{align}
where $\Delta_\Omega$ is the Laplacian in the 2-sphere, namely
\begin{align}
\Delta_\Omega=\frac{1}{\sin\theta}\partial_\theta(\sin\theta\partial_\theta)+\frac{1}{\sin^2\theta}\partial_\varphi^2\,.
\end{align}
Now, plugging Eqs.~\eqref{eq:deltaomega}, \eqref{eq:deltalambda}, \eqref{eq:deltaetaapp} and \eqref{eq:deltarapp} into Eq.~\eqref{eq:deltaNready}, and after some lengthy algebra, we arrive at
\begin{align}
\label{eq:DeltaNliterature}
\Delta_{{\cal N}}(z_s,\hat n)=&\left(2+\frac{{\cal H}'_s}{{\cal H}_s^2}+\frac{2-5s}{r_s{\cal H}_s}-f_{\rm evo}\right)n_iv^i_o\nonumber\\&+b(z)\delta_{n,{\rm c}}+(3-f_{\rm evo}){\cal H}_sv_s+\Psi_s+n_iv^i_s+\frac{1}{{\cal H}_s}\left(\frac{d\Psi_s}{d\lambda_s}-\frac{d}{d\lambda_s}(n_iv^i_s)+\Psi_s'-\Phi_s'\right)\nonumber\\&+\left(5s+\frac{2-5s}{r_s{\cal H}_s}+\frac{{\cal H}'_s}{{\cal H}_s^2}-f_{\rm evo}\right)\left(\Psi_s-n_iv_s^i+\int_0^{\lambda_s}d\lambda\frac{\partial}{\partial \eta}(\Psi-\Phi)\right)\nonumber\\&
+\left(2-5s\right)\Phi_s-\frac{2-5s}{2r_s}\int_0^{\lambda_s}d\lambda\frac{r_s-r}{r}\Delta_\Omega(\Psi-\Phi)
+\frac{2-5s}{r_s}\int_{0}^{\lambda_s}d\lambda(\Psi-\Phi)\,,
\end{align}
where we dropped the monopole terms, and have defined
\begin{align}
\label{eq:sandfevoapp}
s(z)\equiv-\frac{2}{5}\frac{\partial \ln {\cal N}_s}{\partial \ln L}\quad {\rm and}\quad
f_{\rm evo}\equiv -\frac{\partial \ln {\cal N}_s}{\partial \ln (1+z_s)}\,.
\end{align}
Eq.~\eqref{eq:DeltaNliterature} coincides exactly with that given by Challinor \& Lewis \cite{Challinor:2011bk} and Bonvin \& Durrer \cite{Bonvin:2011bg} (see also Ref.~\cite{Nadolny:2021hti}). A useful text  for such calculations is the one by Durrer \cite{Durrer:2020fza}.

\section{Einstein equations \label{app:einstein}}

In this Appendix, we write for completeness the Einstein equations used in the main text. We take the metric to be given by
\begin{align}
d\hat s^2=a^2(\eta)\left(-(1+2\Psi)d\eta^2+(1+2\Phi)\delta_{ij}dx^idx^j\right)\,.
\end{align}
We consider an energy-momentum tensor that includes radiation $\rm r$, pressureless matter $\rm m$ and a cosmological constant $\Lambda$ with
\begin{align}
T^X_{\mu\nu}=(\rho_X+P_X)\hat u_{X\mu} \hat u_{X\nu}+P_Xg_{\mu\nu}\,,
\end{align}
where $X=\{{\rm r},{\rm m},\Lambda\}$. We then have that $w_X\equiv P_X/\rho_X$ is given by $w_{\rm r}=1/3$, $w_{\rm m}=0$, $w_\Lambda=-1$. 

\subsection{Background\label{subsec:background}}

At the background level we have the FL equations which read
\begin{align}
&3M_{\rm pl}^2{\cal H}^2=a^2\sum_X\rho_X\,,\\
&2M_{\rm pl}^2({\cal H}^2-2{\cal H}')=a^2\sum_X\rho_X(1+w_X)\,,
\end{align}
where a prime denotes derivative w.r.t. conformal time, i.e. ${\cal H}'=\partial H/\partial\eta$.
Energy conservation, \textit{i.e.} $\rho_X'+3{\cal H}(1+w_X)\rho_X=0$, then yields $\rho_{\rm m}\propto a^{-3}$ and $\rho_{\rm r}\propto a^{-4}$. For convenience, we define the energy density ratios as
\begin{align}
\Omega_{X}\equiv\frac{a^2\rho_{X}}{3M_{\rm pl}^2{\cal H}^2}\,.
\end{align}
There are analytical solutions available when $\Lambda=0$ and $\rho_r=0$ which are presented below. We set $a_0=1$ in what follows.

\paragraph{Radiation+matter universe:}
The solution to the background equations is:
\begin{align}
\frac{a(\eta)}{a_{\rm eq}}=2\left(\frac{\eta}{\eta_*}\right)+\left(\frac{\eta}{\eta_*}\right)^2
\end{align}
where $(\sqrt{2}-1)\eta_*=\eta_{\rm eq}$ and the subscript ``eq'' refers to the epoch of radiation-matter equality.

\paragraph{Matter+$\Lambda$ universe:}
First, the Hubble parameter is given by
\begin{align}
{\cal H}={\cal H}_0\,a\sqrt{\Omega_{\Lambda,0}+\Omega_{{\rm m},0}a^{-3}}\,.
\end{align}
The conformal time can be written in terms of Hypergeometric functions $F[a,b,c,x]$ as
\begin{align}
\eta=\int_{0}^a \frac{d a_1}{ a_1}\frac{1}{{\cal H}(a_1)}=\frac{2\sqrt{a}}{{\cal H}_0\sqrt{\Omega_{{\rm m},0}}}\,F\left[\tfrac{1}{6},\tfrac{1}{2},\tfrac{7}{6},-\frac{a^3\Omega_{\Lambda,0}}{\Omega_{{\rm m},0}}\right]\,,
\end{align}
In this case, we have an implicit equation for $a(\eta)$.

\subsection{Perturbations \label{app:perturbations}}

To study superhorizon perturbations it is convenient to define the isocurvature perturbation and the relative velocity:
\begin{align}
S\equiv \frac{\delta\rho_{\rm m}}{\rho_{\rm m}}-\frac{3}{4}\frac{\delta\rho_{\rm r}}{\rho_{\rm r}}\quad{\rm and}\quad v_{\rm rel}\equiv v_{\rm m}-v_{\rm r}\,.
\end{align}
With these variables, the linear equations for cosmological perturbations, in Fourier modes, reads:
\begin{align}\label{eq:eomPhi}
\Phi''+3{\cal H}(1+c_s^2)\Phi'+({\cal H}^2(1+3c_s^2)+2{\cal H}')\Phi+c_s^2k^2\Phi=\frac{a^2}{2}\rho_{\rm m}c_s^2S\,,
\end{align}
and
\begin{align}\label{eq:eomS}
S''&+ 3{\cal H}c_s^2S'-\frac{3}{2a^2\rho_{\rm r}}c_s^2{k^4\Phi}+\frac{3\rho_{\rm m}}{4\rho_{\rm r}}c_s^2k^2 S=0\,,
\end{align}
where we also defined 
\begin{align}
c_s^2\equiv\frac{4}{9}\frac{\rho_{\rm r}}{\rho_{\rm m}+4\rho_{\rm r}/3}\,.
\end{align}
Once a solution for $\Phi$ and $S$ is found, the relative velocity is given by $v_{\rm rel}=S'/k^2$. However, for superhorizon modes, it is more convenient to solve the differential equation for $v_{\rm rel}$ which reads
\begin{align}\label{eq:eomsvrel}
v'_{\rm rel}&+3c_s^2{\cal H}v_{\rm rel}+\frac{3}{2a^2\rho_r}c_s^2{\Delta\Phi}+\frac{3\rho_m}{4\rho_r}c_s^2S=0\,.
\end{align}
It is also useful to present the 00-component of Einstein's equations, which is given by
\begin{align}\label{eq:00EE}
&6{\cal H}\Phi'+6{\cal H}^2\Phi-2\Delta\Phi=a^2(\delta\rho_{\rm m}+\delta\rho_{\rm r})\,.
\end{align}

With the solutions to $\Phi$, $S$ and $v_{\rm rel}$, we can follow the evolution of the fluctuations in the matter and radiation energy densities and velocities. With the definitions (see \S~\ref{app:perturbations}) of $\delta_{{\rm m}c}$, $\delta_{{\rm r}c}$, $v_{{\rm m}}$ and $v_{{\rm r}}$, we have that they obey
\begin{align}\label{eq:deltamceq}
&\delta_{{\rm m}c}'+\Delta v_{\rm m}-2a^2\rho_{\rm r}v_{\rm rel}=0\,,\\\label{eq:deltarceq}
&\delta_{{\rm r}c}'-{\cal H}\delta_{{\rm r}c}+\frac{4}{3}\Delta v_{\rm r}+2a^2\rho_{\rm m}v_{\rm rel}=0\,,\\\label{eq:vmeq}
&v_{\rm m}'+{\cal H}v_{\rm m}-\Phi=0\,,\\\label{eq:vreq}
&v_{\rm r}'+{\cal H}v_{\rm r}+\frac{1}{4}\delta_{{\rm r}c}-\Phi=0\,.
\end{align}
We present the analytical solutions below.

\paragraph{Radiation+matter universe:} Here we only focus on superhorizon modes, namely $k\ll{\cal H}$, and so we neglect any $k$ dependence in the equations for perturbations. We may set $\Lambda=0$ as it is negligible at high redshifts. We take as initial conditions $\Phi(a\ll a_{\rm eq})=\Phi_i$ and $S(a\ll a_{\rm eq})=S_i$ and define $\xi\equiv a/a_{\rm eq}$. Then, the solutions for $k\to0$ of \eqref{eq:eomPhi}, \eqref{eq:eomS} and \eqref{eq:eomsvrel} are given by
\begin{align}\label{eq:Sxi}
S(\xi)=&S_i\,,\\
\Phi(\xi)=&\Phi_i \left(\frac{8}{5\xi^3}\left(
\sqrt{1+\xi}-1\right)
-\frac{4}{5\xi^2}
+\frac{1}{5\xi}+\frac{9}{10}\right)\nonumber\\&
+S_i\left(\frac{16}{5\xi^3}
   \left(1-\sqrt{1+\xi}\right)+\frac{8}{5 \xi^2}
   -\frac{2}{5\xi}+\frac{1}{5}\right) \,,\\
v_{\rm rel}(\xi)&=S_i\frac{2 \sqrt{2}}{3 \xi {\cal H}_{\rm eq}} \left(\xi \left(3-\sqrt{1+\xi}\right)+4
   \left(1-\sqrt{1+\xi}\right)\right)\,.
\end{align} 
Integrating \eqref{eq:vmeq} and \eqref{eq:deltamceq} we arrive at
\begin{align}
v_{\rm m}(\xi)=&\sqrt{2}\frac{\Phi_i}{5{\cal H}_{\rm eq}} \left(3  \sqrt{\xi+1}-\frac{4 }{\xi}\frac{\xi-1+\sqrt{1+\xi}}{1+\sqrt{1+\xi}}\right)\nonumber\\&
   +2 \sqrt{2}\frac{S_i}{15{\cal H}_{\rm eq}}\left(
   \sqrt{1+\xi}-\frac{4}{ \xi}\frac{2
   \xi+3-3 \sqrt{1+\xi}}{1+\sqrt{1+\xi}}\right)\,,\\
\delta_{\rm mc}(\xi)=&S_i+S_i\frac{\left(8-\sqrt{1+\xi}+4\xi-\xi^2\right)}{\xi^2}\,.
\end{align}
Lastly, we have: 
\begin{align}\label{eq:vrxi}
v_{\rm r}(\xi)=v_{\rm m}(\xi)-v_{\rm rel}(\xi)\quad{\rm and}\quad
\delta_{\rm rc}(\xi)=\frac{4}{3}\left(\delta_{\rm mc}(\xi)-S_i\right)\,.
\end{align}

\paragraph{Matter+$\Lambda$ universe:}
Here we are interested in low enough redshift such that radiation does not play any important role. Then, the solution to Eq.~\eqref{eq:eomPhi} is \cite{Erickcek:2008jp}
\begin{align}\label{eq:Phia}
\Phi(a)=A\frac{{\cal H}/{\cal H}_0}{a^2}\int_0^{a}\frac{d a_1}{({\cal H}(a_1)/{\cal H}_0)^3}\quad{\rm with}\quad
A=\frac{5}{2}\Omega_{{\rm m},0}\Phi_{\rm MD}\,,
\end{align}
so that the initial conditions are given by $\Phi(a\ll1)=\Phi_{\rm MD}$. With the solution for $\Phi$ and the first integral of \eqref{eq:vmeq} we find
\begin{align}\label{eq:va}
v(a)=\frac{1}{a}\int_0^a da_1 \frac{\Phi(a_1)}{{\cal H}(a_1)}=\frac{A}{a^2{\cal H}_0}\int_0^a d  a_1 \frac{a- a_1}{ a_1}\frac{1}{({\cal H}( a_1)/{\cal H}_0)^3}\,.
\end{align}
where in the second step we used integration by parts as in Appendix~\ref{app:usefulformulas}. Both integrals may be written in terms of hypergeometric functions as 
\begin{align}\label{eq:Phia2}
\Phi(a)=\Phi_{q,\rm MD}F\left[\tfrac{1}{3},1,\tfrac{11}{6},-a^3\frac{\Omega_{\Lambda,0}}{\Omega_{{\rm m},0}}\right]\,.
\end{align}
and
\begin{align}\label{eq:va2}
v(a)=\frac{\Phi_{q,\rm MD}}{3{\cal H}}\left(5-3F\left[\tfrac{1}{3},1,\tfrac{11}{6},-a^3\frac{\Omega_{\Lambda,0}}{\Omega_{{\rm m},0}}\right]\right)\,.
\end{align}
The density fluctuations of CDM start at leading order in $q^2$ and are found from \eqref{eq:deltarceq} to be given by
\begin{align}\label{eq:deltamc2}
\delta_{{\rm m}c}=q^2\int_0^a\frac{da}{a{\cal H}}v_{\rm m}(a)\,.
\end{align}

Some useful relations are for the calculations of the dipole are:
\begin{align}\label{eq:ISWd}
&\int_0^{\lambda} d\lambda_1\, \lambda_1\Phi'(a_1)=-\lambda\Phi(a)+\int_0^{\lambda} d\lambda_1 \Phi(a_1)\,,\\\label{eq:ISW2}
&\int_0^{\lambda} d\lambda_1 \Phi(a)=-\frac{1}{2a^2{\cal H}_0}\int_0^a da_1\frac{a^2-a_1^2}{a_1^2}\frac{1}{ ({\cal H}(a_1)/{\cal H}_0)^3}+\frac{1}{2{\cal H}_0}\int_0^1 da_1\frac{1-a_1^2}{a_1^2}\frac{1}{({\cal H}(a_1)/{\cal H}_0)^3}\,,\\\label{eq:ISW3}
&\int_0^{a} \frac{da_1}{a_1^2({\cal H}(a_1)/{\cal H}_0)^3}=\frac{2{\cal H}_0}{3\Omega_{{\rm m},0}}\left(\frac{1}{{\cal H}}+\eta\right)\,,\\\label{eq:ISW4}
&\int_0^{a} \frac{da_1}{a_1({\cal H}(a_1)/{\cal H}_0)^3}=\frac{2a{\cal H}_0}{3\Omega_{{\rm m},0}{\cal H}(a)}\,.
\end{align}

\section{Useful formulas \label{app:usefulformulas}}
We convert some double definite integrals that appear in our calculations in the main text into single definite integrals by using
\begin{align}
\frac{d}{d\lambda_1}\left(\int_0^{\lambda_1}d\lambda_2({\lambda_1-\lambda_2})Q(\lambda_2)\right)=\int_0^{\lambda_1}d\lambda_2 \,Q(\lambda_2)\,,
\end{align}
\begin{align}
\frac{d}{d\lambda_1}\left(\frac{1}{\lambda_1}\int_0^{\lambda_1}d\lambda_2\frac{\lambda_1-\lambda_2}{\lambda_2}Q(\lambda_2)\right)=\frac{1}{\lambda_1^2}\int_0^{\lambda_1}d\lambda_2 \,Q(\lambda_2)\,,
\end{align}
\begin{align}
\frac{d}{d\lambda_1}\left(\frac{1}{\lambda_1}\int_0^{\lambda_1}d\lambda_2 (\lambda_1-\lambda_2)Q(\lambda_2)\right)=\frac{1}{\lambda_1^2}\int_0^{\lambda_1}d\lambda_2 \lambda_2\,Q(\lambda_2)\,,
\end{align}
\begin{align}
\frac{d}{d\lambda_1}\left(\frac{1}{\lambda_1}\int_0^{\lambda_1}d\lambda_2 (\lambda_1-\lambda_2)\lambda_2Q(\lambda_2)\right)=\frac{1}{\lambda_1^2}\int_0^{\lambda_1}d\lambda_2 \lambda_2^2\,Q(\lambda_2)\,,
\end{align}
where $Q$ may be any variable. For example, we have that after some integration by parts, 
\begin{align}
\int_0^{\lambda}\frac{d\lambda_1}{\lambda_1^2}\int_0^{\lambda_1}d{\lambda_2} \,\lambda_2^2\frac{d^2}{d\lambda_2^2}\Phi(\lambda_2)=\Phi(\lambda)-\frac{2}{\lambda}\int_0^{\lambda_s}{d\lambda_1}\Phi(\lambda_1)\,.
\end{align}

\bibliography{bibliography_dipole.bib} 

\end{document}